\documentclass[twocolumn,aps,superscriptaddress,showpacs,nofootinbib,floatfix]{revtex4-1}
\usepackage{epsfig,bm,feynmf}
\usepackage{graphicx}
\usepackage{amsmath}
\usepackage{dcolumn}

\usepackage{graphicx}
\usepackage[usenames,dvipsnames,svgnames,table]{xcolor}

\usepackage{tikz-feynman,contour}
\tikzfeynmanset{compat=1.0.0}
\usepackage[normalem]{ulem}  

\renewcommand\sout{\bgroup \color{red} \ULdepth=-.5ex \ULset}

\begin{document}


\title{Soft gluon emission from heavy quark scattering in strongly interacting quark-gluon plasma}


\author{Taesoo Song}\email{T.Song@gsi.de}
\affiliation{GSI Helmholtzzentrum f\"{u}r Schwerionenforschung GmbH, Planckstrasse 1, 64291 Darmstadt, Germany}

\author{Ilia Grishmanovskii}
\email{grishm@itp.uni-frankfurt.de}
\affiliation{Institut f\"ur Theoretische Physik, Johann Wolfgang Goethe-Universit\"at, Max-von-Laue-Str.\ 1, D-60438 Frankfurt am Main, Germany}

\author{Olga Soloveva}
\email{soloveva@itp.uni-frankfurt.de}
\affiliation{Helmholtz Research Academy Hesse for FAIR (HFHF), GSI Helmholtz Center for Heavy Ion Physics, Campus Frankfurt, 60438 Frankfurt, Germany}
\affiliation{Institut f\"ur Theoretische Physik, Johann Wolfgang Goethe-Universit\"at, Max-von-Laue-Str.\ 1, D-60438 Frankfurt am Main, Germany}


\begin{abstract}
We apply the Low's theorem to soft gluon emission from  heavy quark scattering in the nonperturbative strongly interacting quark-gluon plasma (sQGP). The sQGP is described in terms of the dynamical quasi-particles 
and adjusted to reproduce the EoS from lQCD at finite temperature and chemical potential. Since the emitted gluon is soft and of long wavelength, it does not provide information on the detailed structure of the scattering, and only the emission from incoming and outgoing partons is enough. It simplifies the calculations making the scattering amplitude factorizable into the elastic scattering and the emission of soft gluon. Imposing a proper upper limit on the emitted gluon energy, we obtain the guage-invariant scattering cross sections of heavy quarks with the massive  
partons of the medium as well as their transport coefficients (momentum drag and diffusion) in the QGP and compare with those from the elastic scattering without gluon emission.
\end{abstract}


\maketitle

\section{Introduction}

Heavy flavor is one of the important probes for the properties of the quark-gluon plasma (QGP) produced in ultra-relativistic heavy-ion collisions~\cite{Uphoff:2012gb,He:2014cla,Cao:2013ita,Gossiaux:2010yx,Das:2015ana,Song:2015sfa,Song:2015ykw,Plumari:2017ntm}.
The production of heavy flavor is reliably described by perturbative Quantum Chromodynamics (pQCD), since a large energy-momentum transfer is required.
However, the  hadronization of heavy quark to a heavy meson or heavy baryon is a soft process whose realization depends on model.
If the heavy quark has a large momentum, phenomenological models such as heavy quark fragmentation functions work well~\cite{Peterson:1982ak}.
On the other hand, the hadronization of soft heavy quarks often adopts the coalescence model where the heavy quark combines with an anti-light quark or with a di-quark to form a heavy meson or a heavy baryon, respectively~\cite{Song:2015sfa,Song:2015ykw}.

The production and hadronization processes of heavy flavor are common in p+p and heavy-ion collisions.
The difference between the two collisions is the presence or absence of a hot dense nuclear matter with which the heavy quark interacts and changes energy-momentum.
A heavy quark with a small momentum is shifted towards a larger momentum by collective flows, while the one with a large momentum is suppressed due to energy loss in the QGP.
They are expressed by the nuclear modification factor which is the heavy flavor distribution in heavy-ion collisions scaled by that in p+p collisions and the number of nucleon+nucleon binary collisions.

A heavy quark interacts with matter through elastic scattering and inelastic scattering.
The former brings about the collisional energy loss of heavy quark, while the latter the radiative energy loss because it induces gluon emission. 
The collisional energy loss is dominant at low or intermediate momentum of a heavy quark, which is taken over by the radiative energy loss at high momentum of heavy quark~\cite{Uphoff:2012gb,Cao:2013ita,Aichelin:2013mra}.

The Parton-Hadron-String Dynamics (PHSD) adopts the Dynamical QuasiParticle Model (DQPM) to describe the strongly interacting partonic matter as well as partonic interactions with massive off-shell quasiparticles, contrary to the massless pQCD partons, whose properties are described by the complex self-energies and spectral functions. The real part of self-energy is related to the pole mass and the imaginary part to the spectral width of partons which are taken in the form of the Hard Thermal Loop (HTL) calculations. 
The DQPM is adjusted to reproduce the lattice equation-of-state (EoS) through the strong coupling which depends on temperature and baryon chemical potential~\cite{Cassing:2009vt,Bratkovskaya:2011wp,Voronyuk:2011jd,Linnyk:2013wma,Linnyk:2015rco,Moreau:2019vhw}.
It has been found that the DQPM which is extended to heavy quark interactions in the QGP reproduces the heavy quark transport coefficients from lattice calculations as well as the experimental data on heavy flavor production in heavy-ion collisions~\cite{Berrehrah:2013mua,Berrehrah:2014kba,Song:2015sfa,Song:2015ykw}.
One limitation of the DQPM for heavy quarks is the absence of radiative energy loss.
Though it can be justified at low and intermediate energies of heavy quarks due to the large gluon mass in the DQPM, the radiative energy loss cannot be neglected at large momenta of heavy quarks~\cite{Grishmanovskii:2022tpb}. 

\begin{figure}
  \centering
  \begin{tikzpicture}
    \begin{feynman}
      \vertex(mu) at ( 0, 0.5);
      \vertex(md) at ( 0, -0.5);
      \vertex(ce) at ( 0, 0);
      \vertex (a) at (-1.,1.5) {$p_1$};
      \vertex (b) at (-1.,-1.5) {$p_2$};
      \vertex (c) at (1., 1.5) {$p_3$};
      \vertex (d) at ( 1.,-1.5) {$p_4$};
      \vertex (d1) at ( 0.4,-0.9);
      \vertex (v) at ( 0.5,1);
      \vertex (e) at ( 1.5,0.9) {$q$};
      \diagram* {
        (a) -- [fermion] (mu) -- [fermion] (c),
        (b) -- [fermion] (md) -- [fermion] (d),
        (v) -- [photon] (e),
        (mu) -- [gluon] (md),
        (ce) -- [gluon, bend left] (d1),
      };
    \end{feynman}
  \end{tikzpicture}
\quad
\quad
  \begin{tikzpicture}
    \begin{feynman}
      \vertex[blob] (m) at ( 0, 0) {\contour{white}{}};
      \vertex (a) at (-1.,1.5) {$p_1$};
      \vertex (b) at (-1.,-1.5) {$p_2$};
      \vertex (c) at (1., 1.5) {$p_3$};
      \vertex (d) at ( 1.,-1.5) {$p_4$};
      \vertex (v) at ( 0.6,0.9);
      \vertex (e) at ( 1.5,0.9) {$q$};
      \diagram* {
        (a) -- [fermion] (m) -- [fermion] (c),
        (b) -- [fermion] (m) -- [fermion] (d),
        (v) -- [photon] (e),
      };
    \end{feynman}
  \end{tikzpicture}
  \caption{ According to Refs.~\cite{Low:1958sn,Song:2018wvd} a complicated scattering process on the left hand side is simplified to the right side Feynman diagram, if the emitted photon is soft.}
\label{low}
\end{figure}
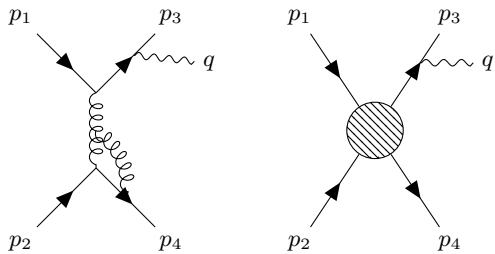

The radiative processes play an important role in quantum electrodynamics (QED). Bremsstrahlung photons are emitted from charged particles which are accelerated or decelerated by scattering (interaction).
According to Refs.~\cite{Low:1958sn,Song:2018wvd} a low-energy photon is emitted from the external charged particles in Feynman diagrams.
In other words, the complicated inner structure of scattering can be ignored in the limit of low energy photon emission, as shown in Fig.~\ref{low}.
Then the Feynman diagram can be factorized into an elastic scattering part and photon emission part.
The reason for ignoring the inner structure of the scattering in the  soft photon limit is found in Ref.~\cite{Song:2018wvd} where the Bremsstrahlung photon spectrum becomes soft with increasing stopping time of the charged particle, while a low-frequency (low-energy) photon is not affected by the stopping time, because the low-frequency photon cannot provide microscopic information on the scattering but only the macroscopic information, for example, incoming and outgoing momenta of the charged particles before and after scattering.

In this study we extend the soft photon approximation to QCD, i.e. to the soft gluon emission from strong interactions.
This extension is reasonable because (anti)quarks and gluons have color charge.
A difference from QED is that the color charge is in SU(3) and non-commutative.
We prove that the soft gluon approximation satisfies the Slavnov-Taylor identities in quark-quark and quark-gluon scatterings, as the soft photon approximation satisfies the Ward-Takahashi identity. 
Then it is applied to the heavy quark scatterings with light quark or gluon with a soft gluon emission in the sQGP as described by the DQPM.

There have been several studies which deal with the soft gluon emission from partonic scatterings~\cite{Gunion:1981qs,Djordjevic:2003zk,Lushozi:2019duv}.
Most of them are focused on gluon emission from the scattering of energetic partons such as jet.
In this case forward scattering is dominant and light-cone coordinate system is convenient, because $\sqrt{s}$ is the largest energy scale, compared to the energy-momentum transfer for the scattering and the emitted gluon energy.
The present study, however, does not assume close to the forward scattering but deals with all possible scattering angles, and treats soft gluon emission systematically up to the leading order of $(q/p)$ with $q$ and $p$ being the energy-momenta of emitted gluon and scattered parton, respectively, such that the Slavnov-Taylor identities are satisfied and the results are explicitly gauge-invariant.

This paper is organized as follows: in section~\ref{photon} the soft photon approximation is rederived up to the leading order for both boson and fermion scatterings, which is extended to QCD in section~\ref{gluon}.
Then the soft gluon emission is applied in section~\ref{QGP} to partonic interactions in the QGP where partons are dressed and thus massive.
The final state phase space  and the scattering cross sections with the soft gluon emission are discussed in section~\ref{cs}, with which the transport coefficients of heavy quarks are calculated in section~\ref{HQ}.
A summary is given in Section~\ref{summary}.

\section{soft photon approximation}\label{photon}

In this section we rederive the formalism for the soft photon emission from both boson and fermion scatterings, which corresponds to the first dominant term in Low's calculations~\cite{Low:1958sn}.

\subsection{Emission from pseudoscalar particles}

\begin{figure}
  \centering
  \begin{tikzpicture}
    \begin{feynman}
      \vertex[blob] (m) at ( 0, 0) {\contour{white}{}};
      \vertex (a) at (-1.,1.5) {$p_1$};
      \vertex (b) at (-1.,-1.5) {$p_2$};
      \vertex (c) at (1., 1.5) {$p_3$};
      \vertex (d) at ( 1.,-1.5) {$p_4$};
      \vertex (v) at ( -0.6,0.9);
      \vertex (e) at (0.2,1.1) {$q$};
      \diagram* {
        (a) -- [fermion] (v) -- [fermion,edge label'=$p_1-q$] (m) -- [fermion] (c),
        (b) -- [fermion] (m) -- [fermion] (d),
        (v) -- [photon] (e),
       };
    \end{feynman}
  \end{tikzpicture}
\quad
\quad
  \begin{tikzpicture}
    \begin{feynman}
      \vertex[blob] (m) at ( 0, 0) {\contour{white}{}};
      \vertex (a) at (-1.,1.5) {$p_1$};
      \vertex (b) at (-1.,-1.5) {$p_2$};
      \vertex (c) at (1., 1.5) {$p_3$};
      \vertex (d) at ( 1.,-1.5) {$p_4$};
      \vertex (v) at ( -0.6,-0.9);
      \vertex (e) at (0.2,-1.1) {$q$};
      \diagram* {
        (a) -- [fermion] (m) -- [fermion] (c),
        (b) -- [fermion] (v) -- [fermion,edge label=$p_2-q$] (m) -- [fermion] (d),
        (v) -- [photon] (e),
       };
    \end{feynman}
  \end{tikzpicture}

  \begin{tikzpicture}
    \begin{feynman}
      \vertex[blob] (m) at ( 0, 0) {\contour{white}{}};
      \vertex (a) at (-1.,1.5) {$p_1$};
      \vertex (b) at (-1.,-1.5) {$p_2$};
      \vertex (c) at (1., 1.5) {$p_3$};
      \vertex (d) at ( 1.,-1.5) {$p_4$};
      \vertex (v) at ( 0.6,0.9);
      \vertex (e) at ( 1.5,0.9) {$q$};
      \diagram* {
        (a) -- [fermion] (m) -- [fermion,edge label'=$p_3+q$] (v) -- [fermion] (c),
        (b) -- [fermion] (m) -- [fermion] (d),
        (v) -- [photon] (e),
      };
    \end{feynman}
  \end{tikzpicture}
\quad
\quad
  \begin{tikzpicture}
    \begin{feynman}
      \vertex[blob] (m) at ( 0, 0) {\contour{white}{}};
      \vertex (a) at (-1.,1.5) {$p_1$};
      \vertex (b) at (-1.,-1.5) {$p_2$};
      \vertex (c) at (1., 1.5) {$p_3$};
      \vertex (d) at ( 1.,-1.5) {$p_4$};
      \vertex (v) at ( 0.6,-0.9);
      \vertex (e) at ( 1.5,-0.9) {$q$};
      \diagram* {
        (a) -- [fermion] (m) -- [fermion] (c),
        (b) -- [fermion] (m) -- [fermion,edge label=$p_4+q$] (v) -- [fermion] (d),
        (v) -- [photon] (e),
      };
    \end{feynman}
  \end{tikzpicture}
  \caption{Photon emission from the four external legs of 2-to-2 scattering}
\label{feynman}
\end{figure}
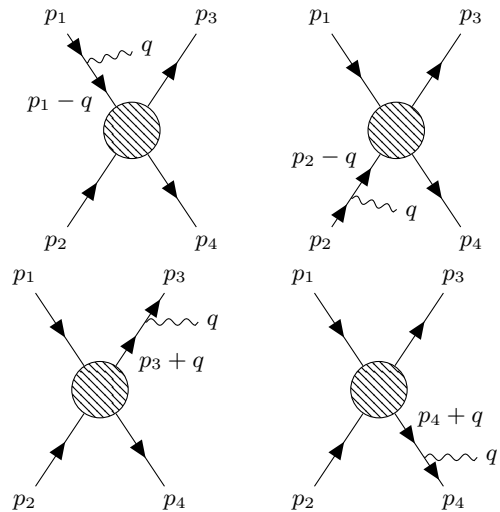

The transition amplitude for photon emission from the scattering of two pseudoscalar particles, as shown in figure~\ref{feynman}, is given by

\begin{eqnarray}
M_{2\rightarrow 2+\gamma}(p_1,p_2;p_3,p_4,q)
=\varepsilon_\mu^*(q)~~~~~~~~~~~~~~~~~~~~~~~~\nonumber\\
\times\{M_{2\rightarrow 2}(p_1-q,p_2;p_3,p_4) G(p_1-q)V^\mu(p_1;p_1-q)\nonumber\\
+M_{2\rightarrow 2}(p_1,p_2-q;p_3,p_4) G(p_2-q)V^\mu(p_2;p_2-q)\nonumber\\
+V^\mu(p_3+q;p_3)G(p_3+q) M_{2\rightarrow 2}(p_1,p_2;p_3+q,p_4)\nonumber\\
+V^\mu(p_4+q;p_4)G(p_4+q) M_{2\rightarrow 2}(p_1,p_2;p_3,p_4+q)\},\nonumber\\
\label{scalar-original}
\end{eqnarray}
where $\varepsilon_\mu^*(q)$ is the polarization vector of the emitted photon, and $G(p)$ and $V^\mu(p+q;p)$ are, respectively, the propagator of the photon-emitting particle and the electromagnetic vertex with photon momentum $q$, which are expressed for the pseudoscalar particle (or pion) as~\cite{Kapusta:1991qp},
\begin{eqnarray}
G(p)&=&\frac{i}{p^2-m^2+i\varepsilon},\nonumber\\
V^\mu(p+q,p)&=&-iQ(2p+q)^\mu.
\end{eqnarray}

Adopting the soft-photon approximation ($q\ll p_1$, $p_2$, $p_3$, $p_4$),
Eq.~(\ref{scalar-original}) is simplified into~\cite{Haglin:1992fy} 
\begin{eqnarray}
&& M_{2\rightarrow 2+\gamma(p_1,p_2;p_3,p_4,q)}\nonumber\\
&&=\varepsilon_\mu^*(q)\bigg\{-\frac{Q_1p_1^\mu}{p_1\cdot q}-\frac{Q_2p_2^\mu}{p_2\cdot q}+\frac{Q_3p_3^\mu}{p_3\cdot q}+\frac{Q_4p_4^\mu}{p_4\cdot q}\bigg\}\nonumber\\
&&~~~\times M_{2\rightarrow 2}(p_1,p_2;p_3,p_4),
\label{scalar}
\end{eqnarray}
which satisfies the Ward-Takahashi identity: $q^\mu (M_{2\rightarrow 2+\gamma})_\mu=$0 where $M_{2\rightarrow 2+\gamma}=\varepsilon^{\mu *}(q)(M_{2\rightarrow 2+\gamma})_\mu$.

\subsection{Emission from fermions}

The photon emission from a fermion is more complicated than the emission from a boson due to the spin of fermion. For example, both propagator and vertex include a gamma matrix:
\begin{eqnarray}
G(p)&=&i\frac{\not{p}+m}{p^2-m^2+i\varepsilon},\nonumber\\
V^\mu(p+q,p)&=&-iQ\gamma^\mu.
\end{eqnarray}

If a photon goes out from $p_3$ as in the lower left diagram of figure~\ref{feynman}, the spinor of $p_3$ is substituted by 
\begin{eqnarray}
\bar{u}^s(p_3)&\rightarrow& -\bar{u}^s(p_3)iQ_3\gamma^\mu i\frac{\not{p_3}+\not{q}+m}{(p_3+q)^2-m^2+i\varepsilon}\varepsilon_\mu^*(q)\nonumber\\
&=&-\bar{u}^s(p_3)iQ_3\gamma^\mu i\frac{u^r(p_3+q)\bar{u}^r(p_3+q)}{(p_3+q)^2-m^2+i\varepsilon}\varepsilon_\mu^*(q)\nonumber\\
&=&Q_3 \frac{\bar{u}^s(p_3)\gamma^\mu u^r(p_3+q)}{2p_3\cdot q}\varepsilon_\mu^*(q)\bar{u}^r(p_3+q).
\label{fermion0}
\end{eqnarray}

Making use of the Gordon decomposition~\cite{Halzen:1984mc},
\begin{eqnarray}
&&\bar{u}^s(p_3)\gamma^\mu u^r(p_3+q)\nonumber\\
&&=\frac{1}{2m}\bar{u}^s(p_3)\{(2p_3+q)^\mu-i\sigma^{\mu\nu}q_\nu\}u^r(p_3+q)\nonumber\\
&&\approx \frac{p_3^\mu}{m}\bar{u}^s(p_3)u^r(p_3)=2p_3^\mu\delta_{sr}
\label{Gordon}
\end{eqnarray}
in the limit $q \ll p_3$, where the superscripts $s$ and $r$ are spin indices and 
\begin{eqnarray}
\sigma^{\mu\nu}=\frac{i}{2}[\gamma^\mu,\gamma^\nu],
\end{eqnarray}
one finds that the modification of the transition amplitude in Eq.~(\ref{fermion0}) is the same as that for a pion in Eq.~(\ref{scalar}):
\begin{eqnarray}
\bar{u}^s(p_3)\rightarrow \varepsilon_\mu^*(q) \frac{Q_3 p_3^\mu}{p_3\cdot q}\bar{u}^s(p_3+q) \approx \varepsilon_\mu^*(q) \frac{Q_3 p_3^\mu}{p_3\cdot q}\bar{u}^s(p_3).
\label{fermion}
\end{eqnarray}
Therefore the soft photon approximation of Eq.~(\ref{scalar}) is applied not only to pseudoscalar particle scattering but also to fermion scattering.

\section{Soft gluon emission}\label{gluon}

Now we apply the same approach to gluon emission, assuming that the emitted gluon is soft and has a long wavelength.

\subsection{Emission from (anti)quarks}

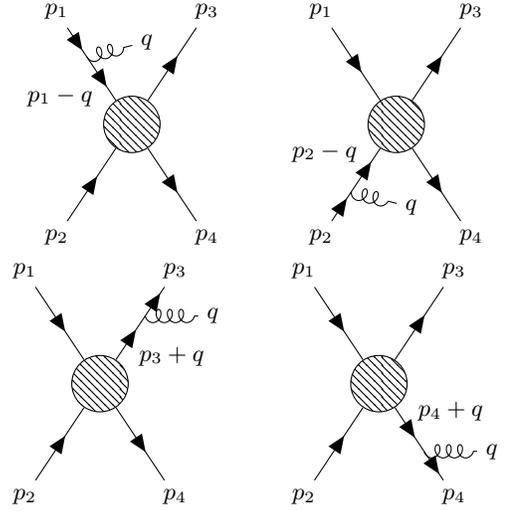
\begin{figure}
  \centering
  \begin{tikzpicture}
    \begin{feynman}
      \vertex[blob] (m) at ( 0, 0) {\contour{white}{}};
      \vertex (a) at (-1.,1.5) {$p_1$};
      \vertex (b) at (-1.,-1.5) {$p_2$};
      \vertex (c) at (1., 1.5) {$p_3$};
      \vertex (d) at ( 1.,-1.5) {$p_4$};
      \vertex (v) at ( -0.6,0.9);
      \vertex (e) at (0.2,1.1) {$q$};
      \diagram* {
        (a) -- [fermion] (v) -- [fermion,edge label'=$p_1-q$] (m) -- [fermion] (c),
        (b) -- [fermion] (m) -- [fermion] (d),
        (v) -- [gluon] (e),
       };
    \end{feynman}
  \end{tikzpicture}
\quad
\quad
  \begin{tikzpicture}
    \begin{feynman}
      \vertex[blob] (m) at ( 0, 0) {\contour{white}{}};
      \vertex (a) at (-1.,1.5) {$p_1$};
      \vertex (b) at (-1.,-1.5) {$p_2$};
      \vertex (c) at (1., 1.5) {$p_3$};
      \vertex (d) at ( 1.,-1.5) {$p_4$};
      \vertex (v) at ( -0.6,-0.9);
      \vertex (e) at (0.2,-1.1) {$q$};
      \diagram* {
        (a) -- [fermion] (m) -- [fermion] (c),
        (b) -- [fermion] (v) -- [fermion,edge label=$p_2-q$] (m) -- [fermion] (d),
        (v) -- [gluon] (e),
       };
    \end{feynman}
  \end{tikzpicture}

  \begin{tikzpicture}
    \begin{feynman}
      \vertex[blob] (m) at ( 0, 0) {\contour{white}{}};
      \vertex (a) at (-1.,1.5) {$p_1$};
      \vertex (b) at (-1.,-1.5) {$p_2$};
      \vertex (c) at (1., 1.5) {$p_3$};
      \vertex (d) at ( 1.,-1.5) {$p_4$};
      \vertex (v) at ( 0.6,0.9);
      \vertex (e) at ( 1.5,0.9) {$q$};
      \diagram* {
        (a) -- [fermion] (m) -- [fermion,edge label'=$p_3+q$] (v) -- [fermion] (c),
        (b) -- [fermion] (m) -- [fermion] (d),
        (v) -- [gluon] (e),
      };
    \end{feynman}
  \end{tikzpicture}
\quad
\quad
  \begin{tikzpicture}
    \begin{feynman}
      \vertex[blob] (m) at ( 0, 0) {\contour{white}{}};
      \vertex (a) at (-1.,1.5) {$p_1$};
      \vertex (b) at (-1.,-1.5) {$p_2$};
      \vertex (c) at (1., 1.5) {$p_3$};
      \vertex (d) at ( 1.,-1.5) {$p_4$};
      \vertex (v) at ( 0.6,-0.9);
      \vertex (e) at ( 1.5,-0.9) {$q$};
      \diagram* {
        (a) -- [fermion] (m) -- [fermion] (c),
        (b) -- [fermion] (m) -- [fermion,edge label=$p_4+q$] (v) -- [fermion] (d),
        (v) -- [gluon] (e),
      };
    \end{feynman}
  \end{tikzpicture}
  \caption{Gluon emission from $q+q\rightarrow q+q$ scattering}
\label{feynman-quark}
\end{figure}

As shown in Fig.~\ref{feynman-quark} gluon emission is the same as photon emission except for a color factor.
The quark propagator and gluon vertex are given by 
\begin{eqnarray}
G_{ij}(p)&=&i\frac{\not{p}+m}{p^2-m^2+i\varepsilon}\delta_{ij},\nonumber\\
V_{ij}^{\mu,a}(p+q,p)&=&ig\gamma^\mu T_{ij}^a,
\end{eqnarray}
where $i,j$ and $a$ are, respectively, the color indices of quark and gluon.
Making the same substitution as in Eq.~(\ref{fermion0}),
\begin{eqnarray}
&&\bar{u}_i^s(p_3)M_{2\rightarrow 2, i}\nonumber\\
&\rightarrow& -g\bar{u}_i^s(p_3)\gamma^\mu T_{ij}^a \varepsilon_\mu^{a*}(q)\frac{\not{p_3}+\not{q}+m}{(p_3+q)^2-m^2+i\varepsilon}M_{2\rightarrow 2, j}\nonumber\\
&=&-g \frac{\bar{u}_i^s(p_3)\gamma^\mu T_{ij}^a u_j^r(p_3+q)}{2p_3\cdot q}\varepsilon_\mu^{a*}(q)\nonumber\\
&&\times\bar{u}_j^r(p_3+q)M_{2\rightarrow 2, j},
\label{quark0}
\end{eqnarray}
one can see that Eq.~(\ref{quark0}) is very similar to Eq.~(\ref{fermion0}) except the color factor $T_{ij}^a$.
The transition amplitude turns out as
\begin{eqnarray}
M_{2q\rightarrow 2q+g}^{kl;ij}(p_1,p_2;p_3,p_4,q)~~~~~~~~~~~~~~~~~~~~~~~~~~~~~\nonumber\\
=g\varepsilon_\mu^{a*}(q)\bigg\{\frac{p_1^\mu}{p_1\cdot q}M_{2q\rightarrow 2q}^{kl;mj}T^a_{mi}+\frac{p_2^\mu}{p_2\cdot q}M_{2q\rightarrow 2q}^{kl;im}T^a_{mj}\nonumber\\
-\frac{p_3^\mu}{p_3\cdot q}T^a_{km}M_{2q\rightarrow 2q}^{ml;ij}-\frac{p_4^\mu}{p_4\cdot q}T^a_{lm}M_{2q\rightarrow 2q}^{km;ij}\bigg\},\nonumber\\
\label{quark}
\end{eqnarray}
where $i,j$ are the color indices of the incoming quarks and $k,l$ those of the outgoing quarks. 

The simplest color structure of $M_{2q\rightarrow 2q}^{kl;ij}$ is from the one-gluon exchange: 
\begin{eqnarray}
M_{2q\rightarrow 2q}^{kl;ij}\sim T^b_{ki} T^b_{lj}=\frac{1}{2}\bigg(\delta_{jk}\delta_{il}-\frac{1}{N_c}\delta_{ik}\delta_{jl}\bigg).
\label{colorq}
\end{eqnarray}
Substituting Eq.~(\ref{colorq}) into Eq.~(\ref{quark}) one finds that the transition amplitude satisfies the Ward-Takahashi identity:
\begin{eqnarray}
q_\mu M_{2q\rightarrow 2q+g}^{\mu, a, kl;ij}(p_1,p_2;p_3,p_4,q)=0,
\label{conservation-q}
\end{eqnarray}
where 
\begin{eqnarray}
M_{2q\rightarrow 2q+g}^{kl;ij}=\varepsilon_\mu^{a*}(q)M_{2\rightarrow 2+g}^{\mu, a, kl;ij}.
\label{m_mu}
\end{eqnarray}
We note that the Ward-Takahashi identity is equivalent to the Slavnov-Taylor identities in the case of on-shell external gluons. 

The transition amplitude squared is given by 
\begin{eqnarray}
|M_{2q\rightarrow 2q+g}|^2=-\frac{g^2}{2N_c}\bigg[
(N_c^2-1)\bigg\{
\frac{m_1^2}{(p_1\cdot q)^2}+\frac{m_2^2}{(p_2\cdot q)^2}\nonumber\\
+\frac{m_3^2}{(p_3\cdot q)^2}+\frac{m_4^2}{(p_4\cdot q)^2}\bigg\}-\frac{4 p_1\cdot p_2}{(p_1\cdot q)(p_2\cdot q)}\nonumber\\
-\frac{4 p_3\cdot p_4}{(p_3\cdot q)(p_4\cdot q)}+\frac{2 p_1\cdot p_3}{(p_1\cdot q)(p_3\cdot q)}+\frac{2 p_2\cdot p_4}{(p_2\cdot q)(p_4\cdot q)}\nonumber\\
-2(N_c^2-2)\bigg\{\frac{p_1\cdot p_4}{(p_1\cdot q)(p_4\cdot q)}+\frac{p_2\cdot p_3}{(p_2\cdot q)(p_3\cdot q)}\bigg\}\bigg]\nonumber\\
\times|M_{2q\rightarrow 2q}|^2.~~~~~
\label{m2-quark}
\end{eqnarray}
The detailed derivations of Eqs.~(\ref{conservation-q}) and (\ref{m2-quark}) are presented in Appendix~\ref{qq-app}.

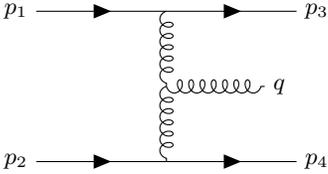
\begin{figure}[ht!]
\centering
\begin{tikzpicture}
  \begin{feynman}
    \vertex (ce) at (0, 0);
    \vertex (v1) at (0, 1);
    \vertex (v2) at (0, -1);
    \vertex (pa) at (-2.,1) {$p_1$};
    \vertex (pb) at (-2.,-1) {$p_2$};
    \vertex (px) at (0,0);
    \vertex (px_) at (-0.6,0) ;
    \vertex (p1) at (2., 1) {$p_3$};
    \vertex (p2) at (2.,-1) {$p_4$};
    \vertex (p3) at (1.5,0) {$q$};
    \diagram* {
      (pa) -- [fermion] (v1) -- [fermion] (p1),
      (px) -- [gluon] (p3),
      (pb) -- [fermion] (v2) -- [fermion] (p2),
      (v1) -- [gluon] (px) -- [gluon] (v2),
      };
  \end{feynman}
\end{tikzpicture}
\caption{Soft gluon emission from the exchanged gluon in quark-quark scattering}
\label{exchange-fig}
\end{figure}

One may think of soft gluon emission from the exchanged gluon as in figure~\ref{exchange-fig}, which is expressed as
\begin{eqnarray}
\varepsilon_\mu^{a*}(q) gf^{abc}[g^{\nu\mu}(-p_1+p_3-2q)^\lambda+g^{\lambda\mu}(-p_1+p_3+q)^\nu\nonumber\\
+g^{\nu\lambda}(2p_1-2p_3+q)^\mu]\frac{-i}{(p_1-p_3+q)^2} \bigg(M_{2q\rightarrow 2q}^{kl;ij}\bigg)_{\nu\lambda}^{bc},~~~
\label{exchange1}
\end{eqnarray}
considering that the one gluon propagator and a three gluon vertex are attached to the original 2-to-2 Feynman diagram.
Since 
\begin{eqnarray}
\bar{u}(p_3)(\not{p_3}-\not{p_1})u(p_1)&=&0,\nonumber\\
\bar{u}(p_4)(\not{p_3}-\not{p_1})u(p_2)&=&\bar{u}(p_4)(\not{p_2}-\not{p_4})u(p_2)=0,\nonumber
\end{eqnarray}
Eq.~(\ref{exchange1}) is simplified in the limit $q\rightarrow 0$ into 
\begin{eqnarray}
-ig\varepsilon_\mu^{a*}(q) \frac{2(p_1-p_3)^\mu}{(p_1-p_3)^2+2(p_1-p_3)\cdot q} f^{abc}\bigg(M_{2q\rightarrow 2q}^{kl;ij}\bigg)^{bc}.\nonumber\\
\label{exchange2}
\end{eqnarray}
Removing $(p_1-p_3)^2$ in the denominator, it looks similar to the terms in Eq.~(\ref{quark}).
Comparing Eqs.~(\ref{quark}) and (\ref{exchange2}) without $M_{2q\rightarrow 2q}$, the former is of the order of $1/q$ and the latter of the order of $1/p=1/q(q/p)$ where $q$ and $p$ are, respectively, the momenta of the soft gluon and of the scattering parton.
Since $q/p \ll 1$, the diagram in figure~\ref{exchange-fig} is of higher order than those in figure~\ref{feynman-quark}, if the transition amplitude is expanded in term of $q/p$:
\begin{eqnarray}
{\rm Fig.~\ref{feynman-quark}}&\sim& \frac{1}{q}\bigg\{A_0+A_1\bigg(\frac{q}{p}\bigg)+...\bigg\},\nonumber\\
{\rm Fig.~\ref{exchange-fig}}&\sim&\frac{1}{q}\bigg\{B_1\bigg(\frac{q}{p}\bigg)+...\bigg\}.
\label{order-counting}
\end{eqnarray}
If the diagram of Fig.~\ref{exchange-fig} is taken into account, one should also include the next-to-leading order term in figure~\ref{feynman-quark}, that is, $A_1(q/p)$ in the above equation, for the final results to be gauge-invariant. 
There is one more important advantage in taking only the leading order in $(q/p)$.
The soft gluon emission from the exchanged virtual gluon is rather simple in $q+q\rightarrow q+q+g$, but much more complicated in $q+g\rightarrow q+g+g$ which will be explained in the next section.
That is why the soft gluon emission is studied only in quark-quark scattering in Ref.~\cite{Gunion:1981qs,Lushozi:2019duv}.
Some works~\cite{Uphoff:2012gb,Aichelin:2013mra} include $q+g\rightarrow q+g+g$ to study the radiative energy loss of heavy quark in QGP.
However, they consider only the gluon emission from t-channel which is dominant in high-energy scattering, though it is not gauge-invariant without $s-$ and $u-$ channels.

In spite of the order counting in Eq.~(\ref{order-counting}), $(p_1-p_3)^2$ in the denominator of Eq.~(\ref{exchange2}) can be smaller than the second term, $(p_1-p_3)\cdot q$, near forward scattering.
Therefore, Eq.~(\ref{quark}) is not a good approximation for nearly forward scattering and the valid kinematic region must properly be restricted. 
It will be discussed in the next section.

\begin{figure}
  \centering
  \begin{tikzpicture}
    \begin{feynman}
      \vertex[blob] (m) at ( 0, 0) {\contour{white}{}};
      \vertex (a) at (-1.,1.5) {$-p_1$};
      \vertex (b) at (-1.,-1.5) {$-p_2$};
      \vertex (c) at (1., 1.5) {$-p_3$};
      \vertex (d) at ( 1.,-1.5) {$-p_4$};
      \vertex (v) at ( -0.6,0.9);
      \vertex (e) at (0.2,1.1) {$q$};
      \diagram* {
        (c) -- [fermion] (m) -- [fermion,edge label=$-p_1+q$] (v) -- [fermion] (a),
        (d) -- [fermion] (m) -- [fermion] (b),
        (v) -- [gluon] (e),
       };
    \end{feynman}
  \end{tikzpicture}
\quad
\quad
  \begin{tikzpicture}
    \begin{feynman}
      \vertex[blob] (m) at ( 0, 0) {\contour{white}{}};
      \vertex (a) at (-1.,1.5) {$-p_1$};
      \vertex (b) at (-1.,-1.5) {$-p_2$};
      \vertex (c) at (1., 1.5) {$-p_3$};
      \vertex (d) at ( 1.,-1.5) {$-p_4$};
      \vertex (v) at ( -0.6,-0.9);
      \vertex (e) at (0.2,-1.1) {$q$};
      \diagram* {
        (c) -- [fermion] (m) -- [fermion] (a),
        (d) -- [fermion] (m) -- [fermion,edge label'=$-p_2+q$] (v) -- [fermion] (b),
        (v) -- [gluon] (e),
       };
    \end{feynman}
  \end{tikzpicture}

  \begin{tikzpicture}
    \begin{feynman}
      \vertex[blob] (m) at ( 0, 0) {\contour{white}{}};
      \vertex (a) at (-1.,1.5) {$-p_1$};
      \vertex (b) at (-1.,-1.5) {$-p_2$};
      \vertex (c) at (1., 1.5) {$-p_3$};
      \vertex (d) at ( 1.,-1.5) {$-p_4$};
      \vertex (v) at ( 0.6,0.9);
      \vertex (e) at ( 1.5,0.9) {$q$};
      \diagram* {
        (c) -- [fermion] (v) -- [fermion,edge label=$-p_3-q$] (m) -- [fermion] (a),
        (d) -- [fermion] (m) -- [fermion] (b),
        (v) -- [gluon] (e),
      };
    \end{feynman}
  \end{tikzpicture}
\quad
\quad
  \begin{tikzpicture}
    \begin{feynman}
      \vertex[blob] (m) at ( 0, 0) {\contour{white}{}};
      \vertex (a) at (-1.,1.5) {$-p_1$};
      \vertex (b) at (-1.,-1.5) {$-p_2$};
      \vertex (c) at (1., 1.5) {$-p_3$};
      \vertex (d) at ( 1.,-1.5) {$-p_4$};
      \vertex (v) at ( 0.6,-0.9);
      \vertex (e) at ( 1.5,-0.9) {$q$};
      \diagram* {
        (c) -- [fermion] (m) -- [fermion] (a),
        (d) -- [fermion] (v) -- [fermion,edge label'=$-p_4-q$] (m) -- [fermion] (b),
        (v) -- [gluon] (e),
      };
    \end{feynman}
  \end{tikzpicture}
  \caption{Gluon emission from $\bar{q}+\bar{q}\rightarrow \bar{q}+\bar{q}$ scattering}
\label{feynman-antiquark}
\end{figure}
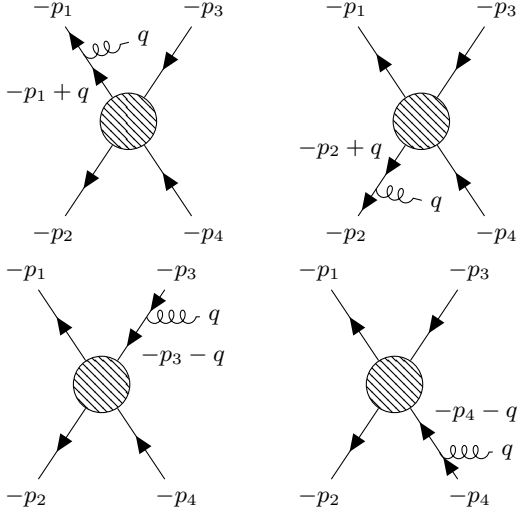

One can think about gluon emission from antiquark scattering ($\bar{q}+\bar{q}\rightarrow \bar{q}+\bar{q}$) as in figure~\ref{feynman-antiquark}.
Then Eq.~(\ref{quark0}) changes into 
\begin{eqnarray}
&&M_{2\bar{q}\rightarrow 2\bar{q}, i}v_i^s(p_3)\nonumber\\
&\rightarrow& g M_{2\rightarrow 2, j}\frac{\not{p_3}+\not{q}-m}{(p_3+q)^2-m^2+i\varepsilon}\varepsilon_\mu^{a*}(q)\gamma^\mu T_{ji}^a v_i^s(p_3)\nonumber\\
&=&g M_{2\rightarrow 2, k}v_k^r(p_3+q)\nonumber\\
&&\times \frac{\bar{v}_j^r(p_3+q) \gamma^\mu T_{ji}^a v_i^s(p_3)}{2p_3\cdot q}\varepsilon_\mu^{a*}(q).
\label{antiquark0}
\end{eqnarray}

The Gordon decomposition for the antifermion current is the same as in Eq.~(\ref{Gordon}) except for an overall minus sign:
\begin{eqnarray}
&&\bar{v}_j^r(p_3+q) \gamma^\mu  v_i^s(p_3)\nonumber\\
&&=-\frac{1}{2m}\bar{v}_j^r(p_3+q) \{(2p_3+q)^\mu+i\sigma^{\mu\nu}q_\nu\}v_i^s(p_3)\nonumber\\
&&\approx -\frac{p_3^\mu}{m}\bar{v}_j^r(p_3) v_i^s(p_3)=2p_3^\mu\delta_{sr}
\label{Gordon2}
\end{eqnarray}
Therefore, the expression of Eq.~(\ref{quark}) is valid not only for $qq$ but also for $\bar{q}\bar{q}$ and $q\bar{q}$ elastic scatterings.


\subsection{Emission from gluon}

\begin{figure}
  \centering
  \begin{tikzpicture}
    \begin{feynman}
      \vertex[blob] (m) at ( 0, 0) {\contour{white}{}};
      \vertex (a) at (-1.,1.5) {$p_1$};
      \vertex (b) at (-1.,-1.5) {$p_2$};
      \vertex (c) at (1., 1.5) {$p_3$};
      \vertex (d) at ( 1.,-1.5) {$p_4$};
      \vertex (v) at ( -0.6,0.9);
      \vertex (e) at (0.2,1.1) {$q$};
      \diagram* {
        (a) -- [fermion] (v) -- [fermion,edge label'=$p_1-q$] (m) -- [fermion] (c),
        (b) -- [gluon] (m) -- [gluon] (d),
        (v) -- [gluon] (e),
       };
    \end{feynman}
  \end{tikzpicture}
\quad
\quad
  \begin{tikzpicture}
    \begin{feynman}
      \vertex[blob] (m) at ( 0, 0) {\contour{white}{}};
      \vertex (a) at (-1.,1.5) {$p_1$};
      \vertex (b) at (-1.,-1.5) {$p_2$};
      \vertex (c) at (1., 1.5) {$p_3$};
      \vertex (d) at ( 1.,-1.5) {$p_4$};
      \vertex (v) at ( -0.6,-0.9);
      \vertex (e) at (0.2,-1.1) {$q$};
      \diagram* {
        (a) -- [fermion] (m) -- [fermion] (c),
        (b) -- [gluon] (v) -- [gluon,edge label=$p_2-q$] (m) -- [gluon] (d),
        (v) -- [gluon] (e),
       };
    \end{feynman}
  \end{tikzpicture}

  \begin{tikzpicture}
    \begin{feynman}
      \vertex[blob] (m) at ( 0, 0) {\contour{white}{}};
      \vertex (a) at (-1.,1.5) {$p_1$};
      \vertex (b) at (-1.,-1.5) {$p_2$};
      \vertex (c) at (1., 1.5) {$p_3$};
      \vertex (d) at ( 1.,-1.5) {$p_4$};
      \vertex (v) at ( 0.6,0.9);
      \vertex (e) at ( 1.5,0.9) {$q$};
      \diagram* {
        (a) -- [fermion] (m) -- [fermion,edge label'=$p_3+q$] (v) -- [fermion] (c),
        (b) -- [gluon] (m) -- [gluon] (d),
        (v) -- [gluon] (e),
      };
    \end{feynman}
  \end{tikzpicture}
\quad
\quad
  \begin{tikzpicture}
    \begin{feynman}
      \vertex[blob] (m) at ( 0, 0) {\contour{white}{}};
      \vertex (a) at (-1.,1.5) {$p_1$};
      \vertex (b) at (-1.,-1.5) {$p_2$};
      \vertex (c) at (1., 1.5) {$p_3$};
      \vertex (d) at ( 1.,-1.5) {$p_4$};
      \vertex (v) at ( 0.6,-0.9);
      \vertex (e) at ( 1.5,-0.9) {$q$};
      \diagram* {
        (a) -- [fermion] (m) -- [fermion] (c),
        (b) -- [gluon] (m) -- [gluon,edge label=$p_4+q$] (v) -- [gluon] (d),
        (v) -- [gluon] (e),
      };
    \end{feynman}
  \end{tikzpicture}
  \caption{Gluon emission from $q+g\rightarrow q+g$ scattering}
\label{feynman-quark-gluon}
\end{figure}
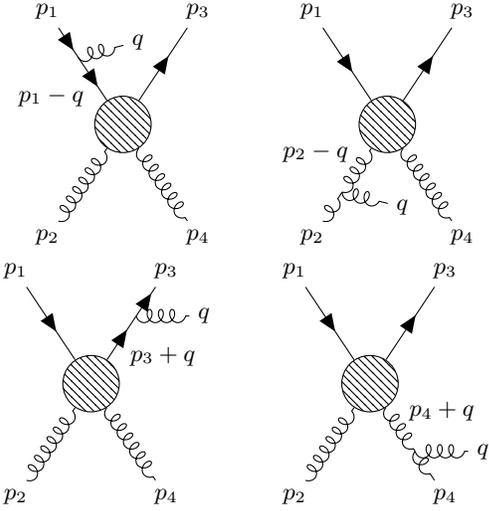

Now we turn to the soft gluon emission from $q+g\rightarrow q+g$ scattering in pQCD as shown in Fig.~\ref{feynman-quark-gluon}.
The gluon propagator and three-gluon vertex are, respectively, given by 
\begin{eqnarray}
G^{\mu\nu,ab}(p)=\frac{-ig^{\mu\nu}}{p^2+i\varepsilon}\delta_{ab},\nonumber\\
V^{\mu\nu\lambda,abc}(p+q,p)=gf^{abc}[g^{\nu\mu}(p+2q)^\lambda\nonumber\\
+g^{\lambda\mu}(-2p-q)^\nu+g^{\nu\lambda}(p-q)^\mu]
\end{eqnarray}
where $\mu,~\nu,~\lambda$ are Lorentz indices and $a,~b,~c$ are color indices of the emitted, incoming and outgoing gluons, respectively.

Making the similar substitution as in Eq.~(\ref{fermion0}),
\begin{eqnarray}
\varepsilon_\lambda^{b*}(p_4)M_{2\rightarrow 2}^{\lambda,b}
\rightarrow -igf^{bdc}\varepsilon_\lambda^{b*}(p_4)\varepsilon_\mu^{c*}(q)\bigg[g^\mu_\nu(p_4+2q)^\lambda\nonumber\\
+g^{\lambda\mu}(p_4-q)_\nu+g_\nu^\lambda(-2p_4-q)^\mu\bigg]\frac{1}{2p_4\cdot q}M_{2\rightarrow 2}^{\nu,d}\nonumber\\
\approx igf^{bdc}\varepsilon_\mu^{c*}(q)\frac{p_4^\mu}{p_4\cdot q}\times\varepsilon_\nu^{b*}(p_4)M_{2\rightarrow 2}^{\nu,d},~~~
\label{gluon4}
\end{eqnarray}
because $p_4^\lambda\varepsilon_\lambda^{c*}(p_4)=0$ and $p_{4\nu}M_{2\rightarrow 2}^{\nu,b}=0$. Similarly,
\begin{eqnarray}
\varepsilon_\lambda^{a}(p_2)M_{2\rightarrow 2}^{\lambda,a}
\rightarrow igf^{adc}\varepsilon_\mu^{c*}(q)\frac{p_2^\mu}{p_2\cdot q}\times\varepsilon_\nu^{a}(p_2)M_{2\rightarrow 2}^{\nu,d}.~~~
\label{gluon2}
\end{eqnarray}

From Eqs.~(\ref{quark}), (\ref{gluon4}) and (\ref{gluon2}) one gets
\begin{eqnarray}
M_{q+g\rightarrow q+g+g}^{\mu,jbc;ia}(p_1,p_2;p_3,p_4,q)~~~~~~~~~~~~~~~~~~~~~~~~~~~~~\nonumber\\
=g\bigg\{\frac{p_1^\mu}{p_1\cdot q}M_{q+g\rightarrow q+g}^{jb;ma}T^c_{mi}
-\frac{p_3^\mu}{p_3\cdot q}T^c_{jm}M_{q+g\rightarrow q+g}^{mb;ia}\nonumber\\
+i\frac{p_2^\mu}{p_2\cdot q}f^{adc}M_{q+g\rightarrow q+g}^{jb;id}+i\frac{p_4^\mu}{p_4\cdot q}f^{bdc}M_{q+g\rightarrow q+g}^{jd;ia}\bigg\},
\label{gluon1}
\end{eqnarray}
where $a$ and $b$ are, respectively, the colors of the incoming and outgoing gluons, and $i$ and $j$ are the colors of the incoming and outgoing quarks, respectively.

Considering that the color structure of $M_{q+g\rightarrow q+g}^{jb;ia}$ is given by $[T^a,T^b]_{ji}$ or $if^{abc}T_{ji}^c$, the matrix element in Eq.~(\ref{gluon1}) satisfies current conservation:
\begin{eqnarray}
q_\mu M_{q+g\rightarrow q+g+g}^{\mu, jbc;ia}(p_1,p_2;p_3,p_4,q)=0,
\label{conservation-g}
\end{eqnarray}
where 
\begin{eqnarray}
M_{q+g\rightarrow q+g+g}^{jb;ia}=\varepsilon_\mu^{c*}(q)M_{q+g\rightarrow q+g+g}^{\mu, jbc;ia},
\label{m_mug}
\end{eqnarray}
and the transition amplitude squared is given by 
\begin{eqnarray}
&&|M_{q+g\rightarrow q+g+g}|^2=-g^2\bigg[\frac{N_c^2-1}{2N_c}\bigg(\frac{m_1^2}{(p_1\cdot q)^2}+\frac{m_3^2}{(p_3\cdot q)^2}\bigg)\nonumber\\
&&+\frac{1}{N_c}\frac{p_1\cdot p_3}{(p_1\cdot q)(p_3\cdot q)}-\frac{N_c}{2}\bigg(\frac{2p_2\cdot p_4}{(p_2\cdot q)(p_4\cdot q)}\nonumber\\
&&+\frac{p_1\cdot p_2}{(p_1\cdot q)(p_2\cdot q)}+\frac{p_3\cdot p_4}{(p_3\cdot q)(p_4\cdot q)}
+\frac{p_1\cdot p_4}{(p_1\cdot q)(p_4\cdot q)}\nonumber\\
&&+\frac{p_2\cdot p_3}{(p_2\cdot q)(p_3\cdot q)}\bigg)\bigg]\times |M_{q+g\rightarrow q+g}|^2.
\label{m2-gluon}
\end{eqnarray}
We note that the transition amplitude, in principle, must be divided by 2, for the two gluons in the final state are indistinguishable.
Since it is assumed that one gluon is the hard gluon involved in the elastic scattering and the other gluon the soft gluon emitted from the scattering, 2 is not divided in Eq.~(\ref{m2-gluon}).
The proof and the derivation are presented in Appendix~\ref{qg-app}.

\section{Soft gluon emission in the sQGP within the DQPM}\label{QGP}

In the Dynamical Quasi-Particle Model (DQPM), quark and gluon are dressed in the sQGP and gain effective masses ($m_{q,g}$) and spectral widths ($\Gamma_{q,g}$) which depend on temperature and quark chemical potential ~\cite{Moreau:2019vhw}. Main properties and parameters of the DQPM is described in Appendix~\ref{dqpm_param}.
The finite width is reflecting the dynamical modification of spectral function of quasiparticles during their propagation in the sQGP medium.
In the DQPM the propagators of quark and gluon are modified, respectively, into
\begin{eqnarray}
G_{ij}(p)&=&i\frac{\not{p}+m_q(T,\mu)}{p^2-m_q^2(T,\mu)+i|p_0|\Gamma_q(T,\mu)}\delta_{ij},\nonumber\\
G^{\mu\nu,ab}(p)&=&\frac{-ig^{\mu\nu}}{p^2-m_g^2(T,\mu)+i|p_0| \Gamma_g(T,\mu)}\delta_{ab}.
\end{eqnarray}

Then the scalar products of the external momentum $p_i$ and the soft gluon momentum $q$ in the denominators of Eq.~(\ref{m2-quark}) are modified as 
\begin{eqnarray}
p_1\cdot q\rightarrow p_1\cdot q-m_g^2/2-i|p_1^0-q^0|\Gamma_q/2,\nonumber\\
p_2\cdot q\rightarrow p_2\cdot q-m_g^2/2-i|p_2^0-q^0|\Gamma_q/2,\nonumber\\
p_3\cdot q\rightarrow p_3\cdot q+m_g^2/2+i|p_3^0+q^0|\Gamma_q/2,\nonumber\\
p_4\cdot q\rightarrow p_4\cdot q+m_g^2/2+i|p_4^0+q^0|\Gamma_q/2,\nonumber
\end{eqnarray}
and in Eq.~(\ref{m2-gluon}) $\Gamma_g(T,\mu)$ is introduced for $p_2\cdot q$ and $p_4\cdot q$:
\begin{eqnarray}
p_2\cdot q\rightarrow p_2\cdot q-m_g^2/2-i|p_2^0-q^0|\Gamma_g/2,\nonumber\\
p_4\cdot q\rightarrow p_4\cdot q+m_g^2/2+i|p_4^0+q^0|\Gamma_g/2.\nonumber
\end{eqnarray}

Since the product is now replaced with a complex function due to the imaginary term, the following substitution must be made:
\begin{eqnarray}
&&\frac{1}{(p_i\cdot q)(p_j\cdot q)}\rightarrow {\rm Re}\bigg[\frac{1}{(p_i\cdot q)(p_j\cdot q)}\bigg]\nonumber\\
&&~~~~=\frac{1}{2}\bigg\{\frac{1}{(p_i\cdot q)(p_j\cdot q)^*}+\frac{1}{(p_i\cdot q)^*(p_j\cdot q)}\bigg\}.
\end{eqnarray}
For example, 
\begin{eqnarray}
&&\frac{1}{(p_1\cdot q)(p_3\cdot q)}\rightarrow \nonumber\\
&&\frac{1}{2}\bigg\{\frac{1}{p_1\cdot q-m_g^2/2-i|p_1^0-q^0|\Gamma_q/2}\nonumber\\
&&~~~\times \frac{1}{p_3\cdot q+m_g^2/2-i|p_3^0+q^0|\Gamma_q/2}\nonumber\\
&&~+\frac{1}{p_1\cdot q-m_g^2/2+i|p_1^0-q^0|\Gamma_q/2}\nonumber\\
&&~~~\times \frac{1}{p_3\cdot q+m_g^2/2+i|p_3^0+q^0|\Gamma_q/2}\bigg\}.
\end{eqnarray}


Furthermore, the first term in the square bracket of Eq.~(\ref{m2-gluon}) has additional terms which vanish for massless gluons:
\begin{eqnarray}
&&-\frac{N_c^2-1}{2N_c}\bigg(\frac{m^2}{(p_1\cdot q)^2}+\frac{m^2}{(p_3\cdot q)^2}\bigg)\rightarrow\nonumber\\
&&-\frac{N_c^2-1}{2N_c}\bigg\{\frac{m_q^2}{\bigg|p_1\cdot q-m_g^2/2-i|p_1^0-q^0|\Gamma_q/2\bigg|^2}\nonumber\\
&&~~~~~~~~~+\frac{m_q^2}{\bigg|p_3\cdot q+m_g^2/2+i|p_3^0+q^0|\Gamma_q/2\bigg|^2}\bigg\}\nonumber\\
&&-N_c\bigg\{\frac{m_g^2}{\bigg|p_2\cdot q-m_g^2/2-i|p_2^0-q^0|\Gamma_g/2\bigg|^2}\nonumber\\
&&~~~~+\frac{m_g^2}{\bigg|p_4\cdot q+m_g^2/2+i|p_4^0+q^0|\Gamma_g/2\bigg|^2}\bigg\}.
\end{eqnarray}

\section{Phase space and cross section}\label{cs}

The cross section for 2-to-3 scattering is given by
\begin{eqnarray}
\sigma_{2\rightarrow 3}=\frac{1}{4p_i \sqrt{s}}\int \frac{d^3p_3}{(2\pi)^32E_3}\int \frac{d^3p_4}{(2\pi)^32E_4}\int \frac{d^3q}{(2\pi)^32E_g}\nonumber\\
\times(2\pi)^4\delta^{(4)}(p_1+p_2-p_3-p_4-q)\overline{|M_{2\rightarrow 3}|}^2,~~~
\end{eqnarray}
where $p_i$ is the initial three momentum in the center-of-mass frame and the line over the transition amplitude squared implies spin-color average.
Introducing the variable $p=p_3+p_4$,
\begin{eqnarray}
\sigma_{2\rightarrow 3}=\frac{1}{4p_i \sqrt{s}}\int d^4p \int \frac{d^3p_3d^3p_4}{(2\pi)^64E_3E_4}\delta^{(4)}(p-p_3-p_4)\nonumber\\
\times\int \frac{d^3q}{(2\pi)^32E_g}(2\pi)^4\delta^{(4)}(p_1+p_2-p-q)\overline{|M_{2\rightarrow 3}|}^2\nonumber\\
=\frac{1}{32\pi p_i \sqrt{s}}\int \frac{d^3q}{(2\pi)^32E_g} \frac{|{\bf p}_3|}{\sqrt{s_2}}\int d\cos\theta\overline{|M_{2\rightarrow 3}}|^2~~~~
\end{eqnarray}
with the constraint of energy-momentum conservation in the second equation. In the above equation $s_2=(p_3+p_4)^2$, and $|{\bf p}_3|$ and $\theta$ are, respectively, the three-momentum and scattering angle of $p_3$ in the center-of-mass frame of $p_3+p_4$:
\begin{eqnarray}
|{\bf p}_3|=\sqrt{\frac{\{s_2-(m_3+m_4)^2\}\{s_2-(m_3-m_4)^2\}}{4 s_2}}.
\end{eqnarray}

Assuming that the emitted gluon is soft, the center-of-mass frame of  $p_3+p_4$ is similar to that of $p_1+p_2$ and the differential cross section is approximated as
\begin{eqnarray}
\frac{d\sigma_{2\rightarrow 3}}{d\cos\theta}
\approx\frac{1}{32\pi p_i \sqrt{s}}\int \frac{d^3q}{(2\pi)^32E_g} \frac{|{\bf p}_3|}{\sqrt{s_2}}\overline{|M_{2\rightarrow 3}}|^2\nonumber\\
\approx \frac{d\sigma_{2\rightarrow 2}}{d\cos\theta}\int \frac{d^3q}{(2\pi)^32E_g} |\epsilon\cdot J|^2\frac{|{\bf p}_3|\sqrt{s}}{p_f\sqrt{s_2}},
\label{approx1}
\end{eqnarray}
where 
\begin{eqnarray}
\overline{|M_{2\rightarrow 3}}|^2&\equiv& |\epsilon\cdot J|^2~ \overline{|M_{2\rightarrow 2}}|^2\nonumber\\
&=&32\pi s|\epsilon\cdot J|^2\frac{p_i}{p_f}\frac{d\sigma_{2\rightarrow 2}}{d\cos\theta}
\label{def1}
\end{eqnarray}
with 
\begin{eqnarray}
p_f=\sqrt{\frac{\{s-(m_3+m_4)^2\}\{s-(m_3-m_4)^2\}}{4 s}}.
\end{eqnarray}
We note that $|{\bf p}_3|\sqrt{s}/(p_f\sqrt{s_2})$ in Eq.~(\ref{approx1}) is responsible for the reduction of phase space of $p_3+p_4$ in 2-to-3 process, compared to 2-to-2 process, because $s_2$ is always smaller than $s$~\cite{Linnyk:2015rco}.

For simplicity we patch gluon energy-momentum to 2-to-2 elastic scattering, ignoring energy-momentum conservation:
\begin{eqnarray}
p_1&=&(E_1,~0,~0,~p_1),\label{p1p4}\\
p_2&=&(E_2,~0,~0,~-p_1),\nonumber\\
p_3&=&(E_3,~0,~p_1\sin\theta, ~p_1\cos\theta),\nonumber\\
p_4&=&(E_4,~0,~-p_1\sin\theta, ~-p_1\cos\theta),\nonumber\\
q&=&(E_g,~{\bf q}\sin\theta'\cos\phi',~{\bf q}\sin\theta'\sin\phi',~{\bf q}\cos\theta'),\nonumber
\end{eqnarray}
which will be used to calculate $|\epsilon\cdot J|^2$ in Eq.~(\ref{def1}). The integration in Eq.~(\ref{approx1}) is then expressed as
\begin{eqnarray}
\int \frac{d^3q}{E_g}=\int_{m_g}^{E_{max}} \sqrt{E_g^2-m_q^2} \ dE_g d\cos\theta' d\phi',
\end{eqnarray}
where $E_{max}$ is the maximum energy of soft gluon. 
From energy conservation it is given by
\begin{eqnarray}
q_{max}^2=\frac{\{s-(m_3+m_4+m_g)^2\}\{s-(m_3+m_4-m_g)^2\}}{4s}\nonumber\\
\label{limit1}
\end{eqnarray}
with $E_{max}=\sqrt{m_g^2+q_{max}^2}$.

However, we have neglected gluon emission from the interaction region which is figured as a blob in the Feynman diagrams, because the wavelength of the emitted gluon is assumed larger than the scattering scale which is roughly $1/\sqrt{-t}$  with $t=(p_1-p_3)^2$.
Therefore, more reasonable limit for our calculations to be valid will be
\begin{eqnarray}
E_{max}={\rm min}\bigg[E_{max}~{\rm from ~Eq.}~(\ref{limit1}),~\sqrt{-t}\bigg],
\label{limit2}
\end{eqnarray} 
which also justifies the ignoring of figure~\ref{exchange-fig}.
From here on the upper limit from Eq.~(\ref{limit1}) will be denoted the kinematic upper limit and Eq.~(\ref{limit2}) the realistic upper limit.

\begin{figure*}[th!]
\centerline{
\includegraphics[width=8.6 cm]{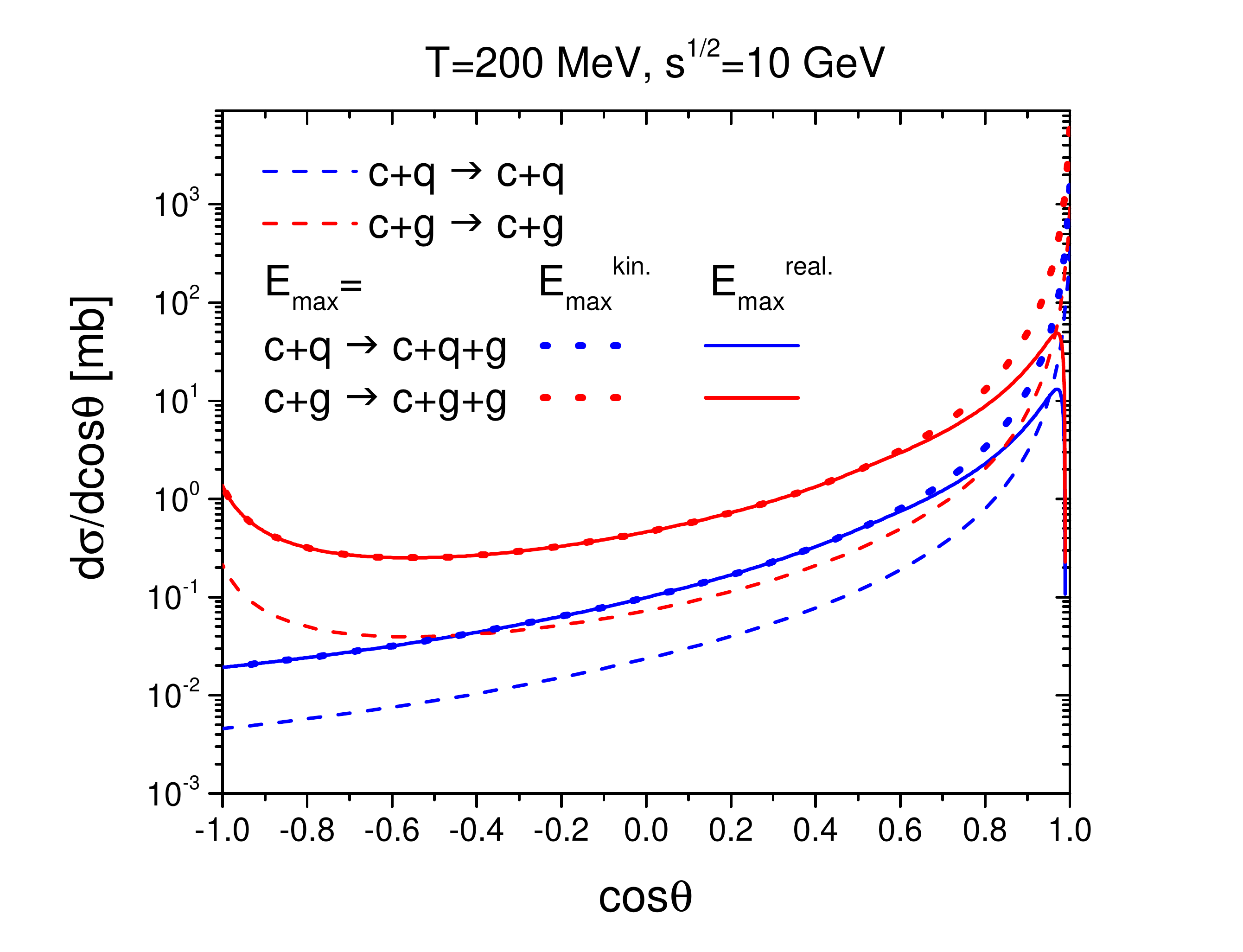}
\includegraphics[width=8.6 cm]{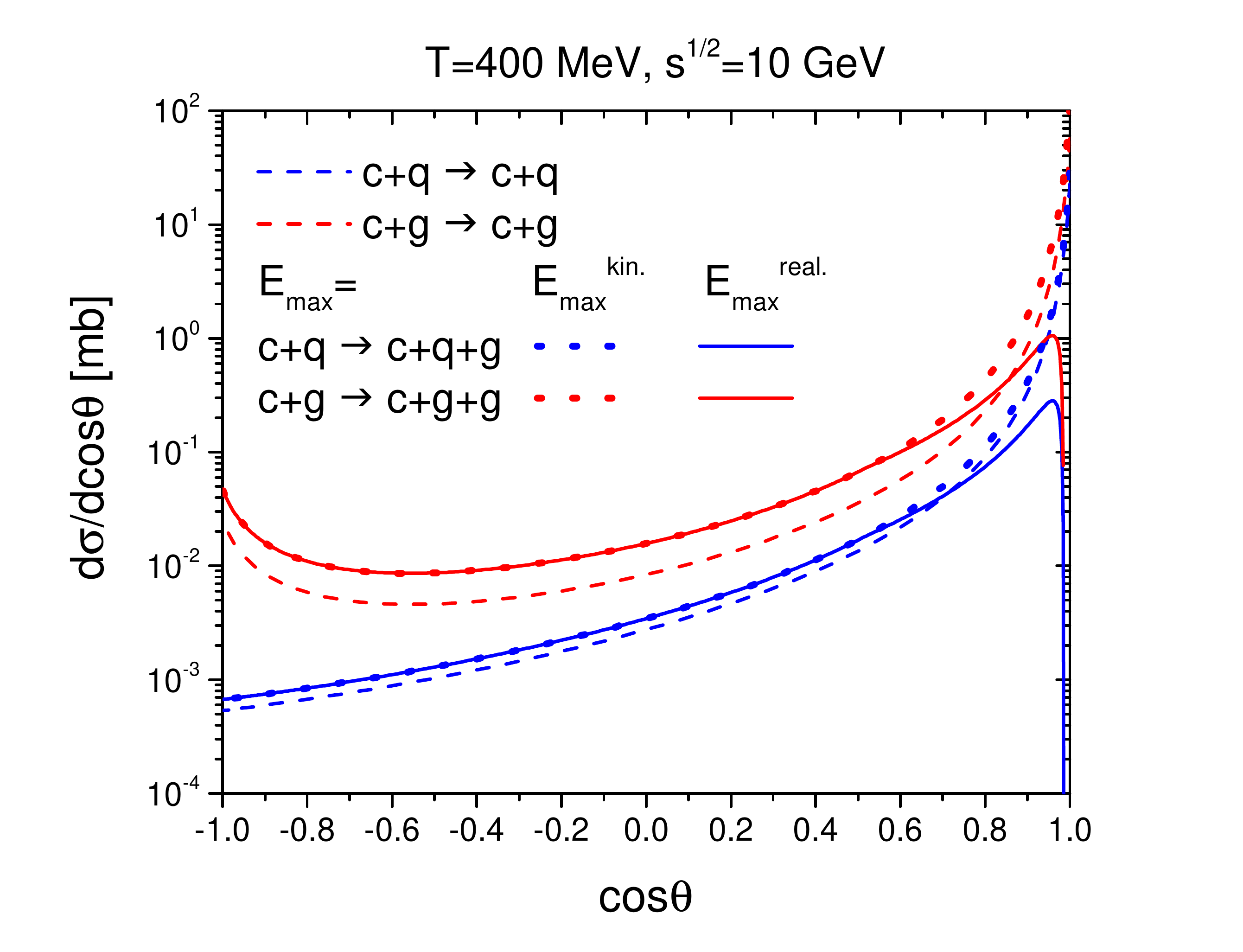}}
\centerline{
\includegraphics[width=8.6 cm]{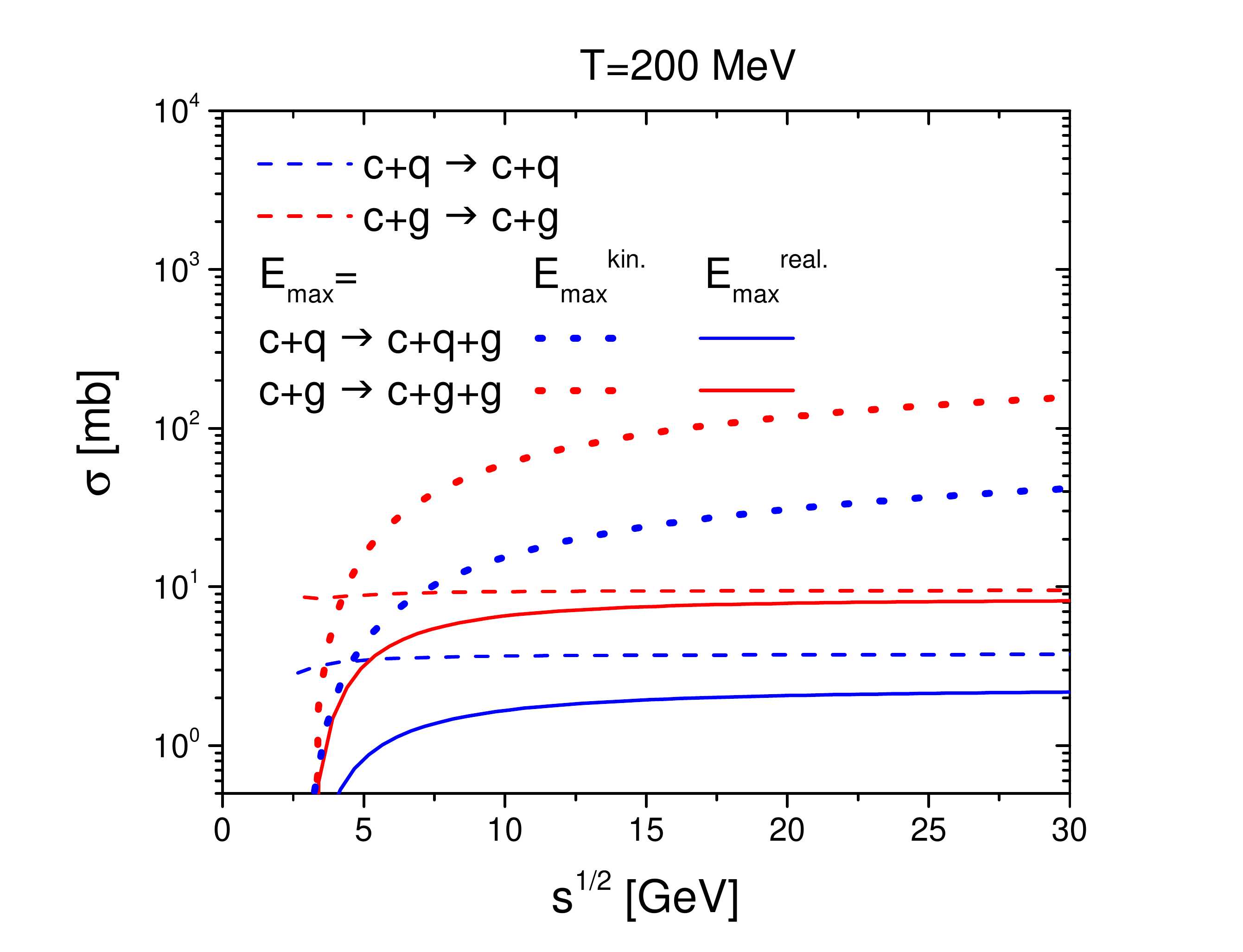}
\includegraphics[width=8.6 cm]{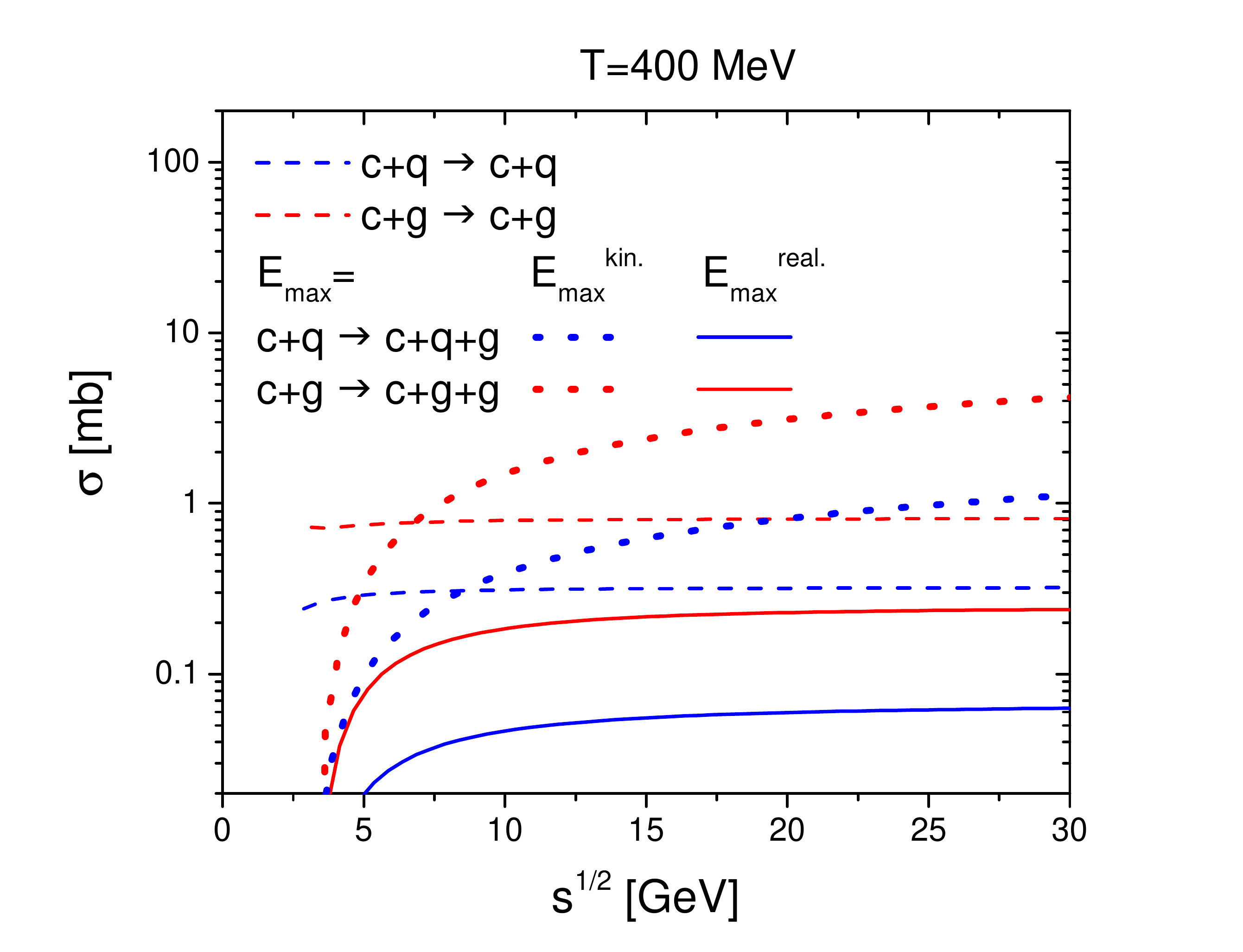}}
\centerline{
\includegraphics[width=8.6 cm]{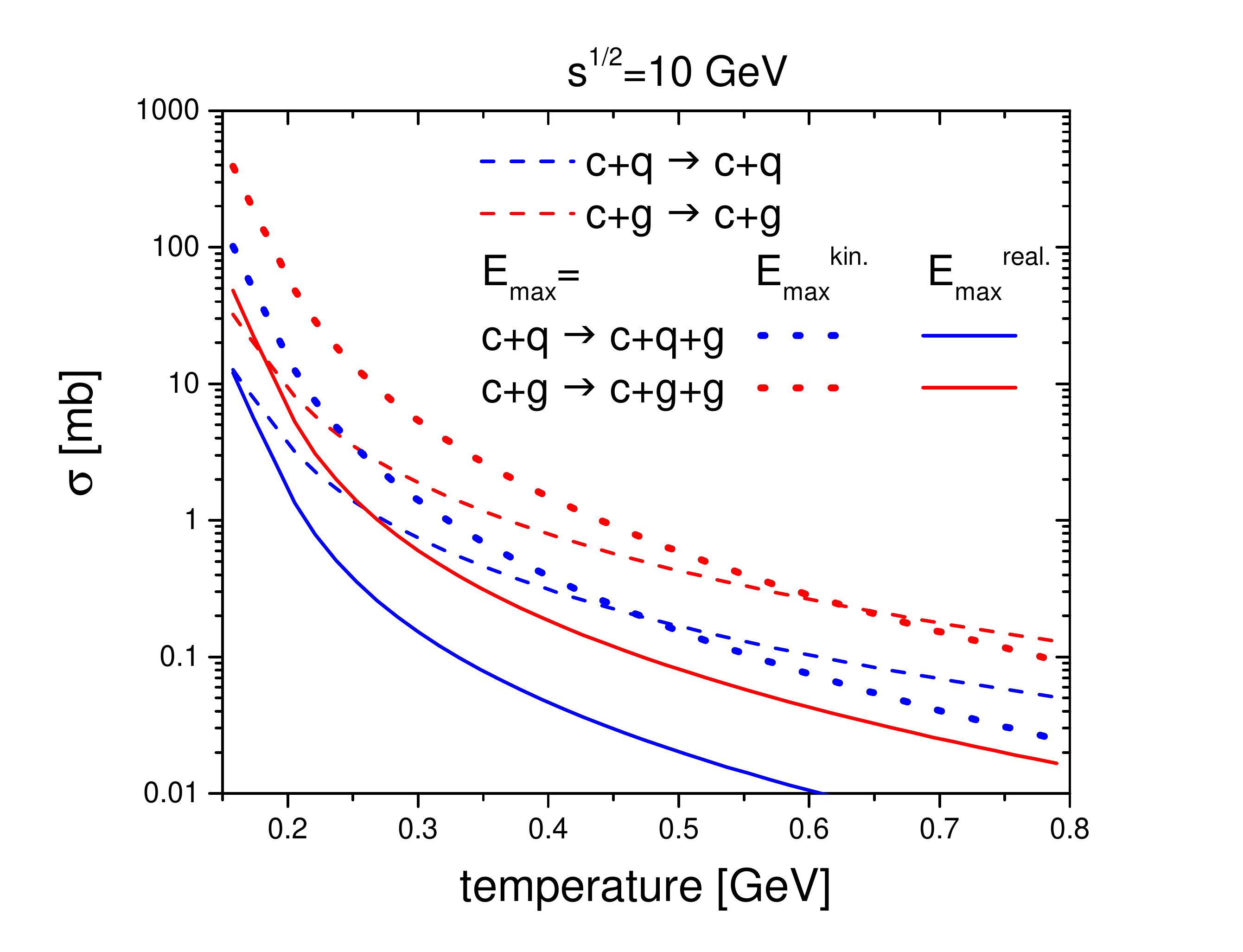}}
\caption{(Color online) Differential and integrated cross sections for $c+q(g)\rightarrow c+q(g)$ and $c+q(g)\rightarrow c+q(g)+g$ with the kinematic and realistic upper limits of soft gluon energy, which are, respectively, displayed by dashed, dotted and solid lines as a function of (upper) the scattering angle of the charm quark, (middle) scattering energy and (lower) temperature.}
\label{sigma}
\end{figure*}

Now we apply the formula, which has been derived, to the soft gluon emission from charm quark elastic scattering in the QGP with charm quark mass being 1.5 GeV.

Figure~\ref{sigma} shows the differential and integrated cross sections for $c+q(g)\rightarrow c+q(g)$ and $c+q(g)\rightarrow c+q(g)+g$ scattering with the kinematic and realistic upper limits of gluon energy, which are, respectively, displayed by dashed, dotted and solid lines, as a function of the scattering angle of the charm quark, scattering energy and temperature.

One can see that the differential cross sections for gluon emission with the two different upper limits of the gluon energy are almost same except near $\cos\theta=1$, i.e. near forward scattering where $t$ is small and thus the realistic upper limit of the soft gluon energy is low.

The middle panels show that the integrated cross sections for gluon emission are suppressed at low energy, because the collision energy is not enough or hard to produce a massive gluon in the final state.
The integrated cross section for the kinematic upper limit is much larger than for the realistic upper limit, since the forward scattering cross section is very large which, however, is suppressed in the latter case.

Finally, comparing 2-to-2 scattering cross sections and 2-to-3 scattering cross sections in the lower panel, the temperature dependence is stronger for the latter.
The reason is that the former is proportional to $\alpha_s^2$ while the latter to $\alpha_s^3$, and $\alpha_s$ is very large near $T_c$.
Therefore, charm quark scattering with gluon emission is more suppressed than the elastic scattering with increasing temperature.

\section{Transport coefficients of heavy quark in the sQGP}\label{HQ}

Charm quarks change their energy-momentum in the QGP with time through scattering, which is expressed in a Fokker-Planck equation as~\cite{Svetitsky:1987gq}:
\begin{eqnarray}
\frac{\partial f(p)}{\partial t}=\frac{\partial}{\partial p_i}\bigg[A_i(p)f(p) +\frac{\partial}{\partial p_i}[B_{ij}(p)f(p)]\bigg],
\end{eqnarray}
where $f(p)$ is the charm distribution function and the drag, transverse/longitudinal diffusion coefficients and $\hat{q}$ of the charm quark in the medium, supposing that the charm quark moves in $z-$direction, are defined as
\begin{eqnarray}
A(p)&=&-\frac{\vec{A}(p)\cdot \vec{p}}{|\vec{p}|}= -\frac{d\langle \Delta p \rangle}{dt}=\langle\langle(p-p^\prime)_z\rangle\rangle,\label{A}\\
B_L(p)&=& \frac{1}{2}\frac{p_ip_j}{|\vec{p}|^2}B_{ij}(p)\nonumber\\
&=&\frac{1}{2}\frac{d\langle (\Delta p_L)^2 \rangle}{dt}= \frac{1}{2}\langle\langle(p-p^\prime)_z^2\rangle\rangle,\label{bldef}\\
B_T(p)&=& \frac{1}{4}\bigg(\delta_{ij}-\frac{p_ip_j}{|\vec{p}|^2}\bigg)B_{ij}(p)\nonumber\\
&=&\frac{1}{4}\frac{d\langle (\Delta p_T)^2 \rangle}{dt}= \frac{1}{4}\langle\langle p_x^{\prime 2}+p_y^{\prime 2}\rangle\rangle,\label{btdef}\\
\hat{q}(p)&=&\frac{d\langle(\Delta p_T)^2\rangle}{d z}=\frac{4E}{p_L}B_T(p),
\label{qhat}
\end{eqnarray}
where $p_L$ and $p_T$ are respectively the longitudinal and transverse momentum of charm quark.

The double bracket in Eqs.~(\ref{A}) to (\ref{qhat}) implies~\cite{Berrehrah:2014kba,Song:2019cqz}
\begin{eqnarray}
\langle\langle O^* \rangle\rangle \equiv \sum_{i=q,\bar{q},g}\int dm_i dm_f dm_g A_i(m_i)A_i(m_f)A_g(m_g)\nonumber\\
\times \int \frac{d^3k}{(2\pi)^3 }f_i(k)~O^*~ v_{ic}\sigma_{ic},~~~~~~~~~~
\label{def2}
\end{eqnarray}
 for the scattering of offshell partons, 
where $m_i$, $m_f$ and $m_g$ are respectively the incoming and outgoing parton masses and the emitted gluon mass, and $A_i$, $A_f$ and $A_g$ their spectral functions whose pole masses and spectral widths are presented in Appendix~\ref{dqpm_param};  $f_i(k)$ is a distribution function of parton $i$, and $v_{ic}$ and $\sigma_{ic}$ are the relative velocity and the scattering cross section of the charm quark and parton $i$, respectively.
We note that the scattering cross section in Eq.~(\ref{def2}) is multiplied by 2 in order to reproduce the lattice data on the spatial diffusion coefficient and the experimental data on heavy flavors in heavy-ion collisions~\cite{Song:2016rzw}.

\begin{figure*}[th!]
\centerline{
\includegraphics[width=8.6 cm]{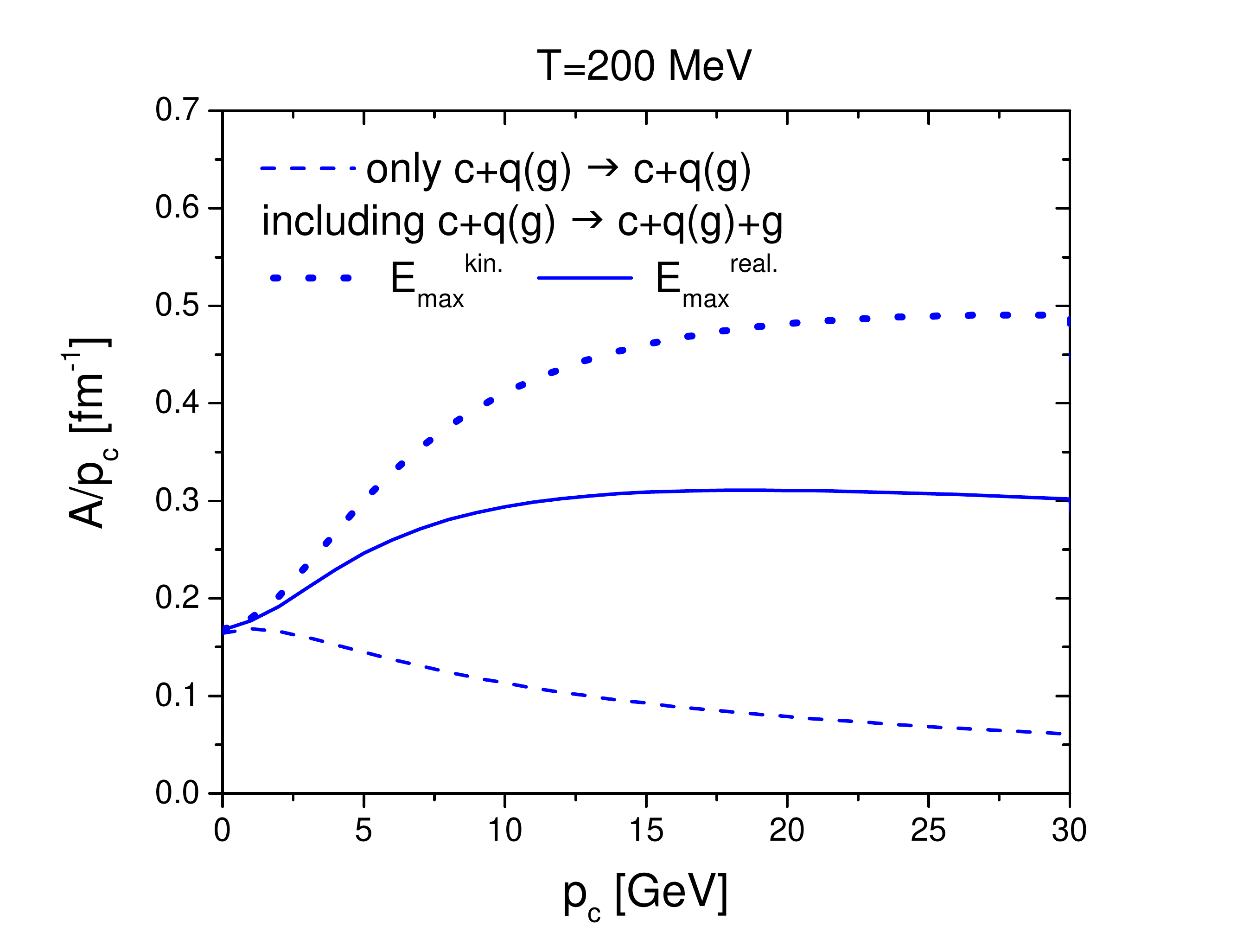}
\includegraphics[width=8.6 cm]{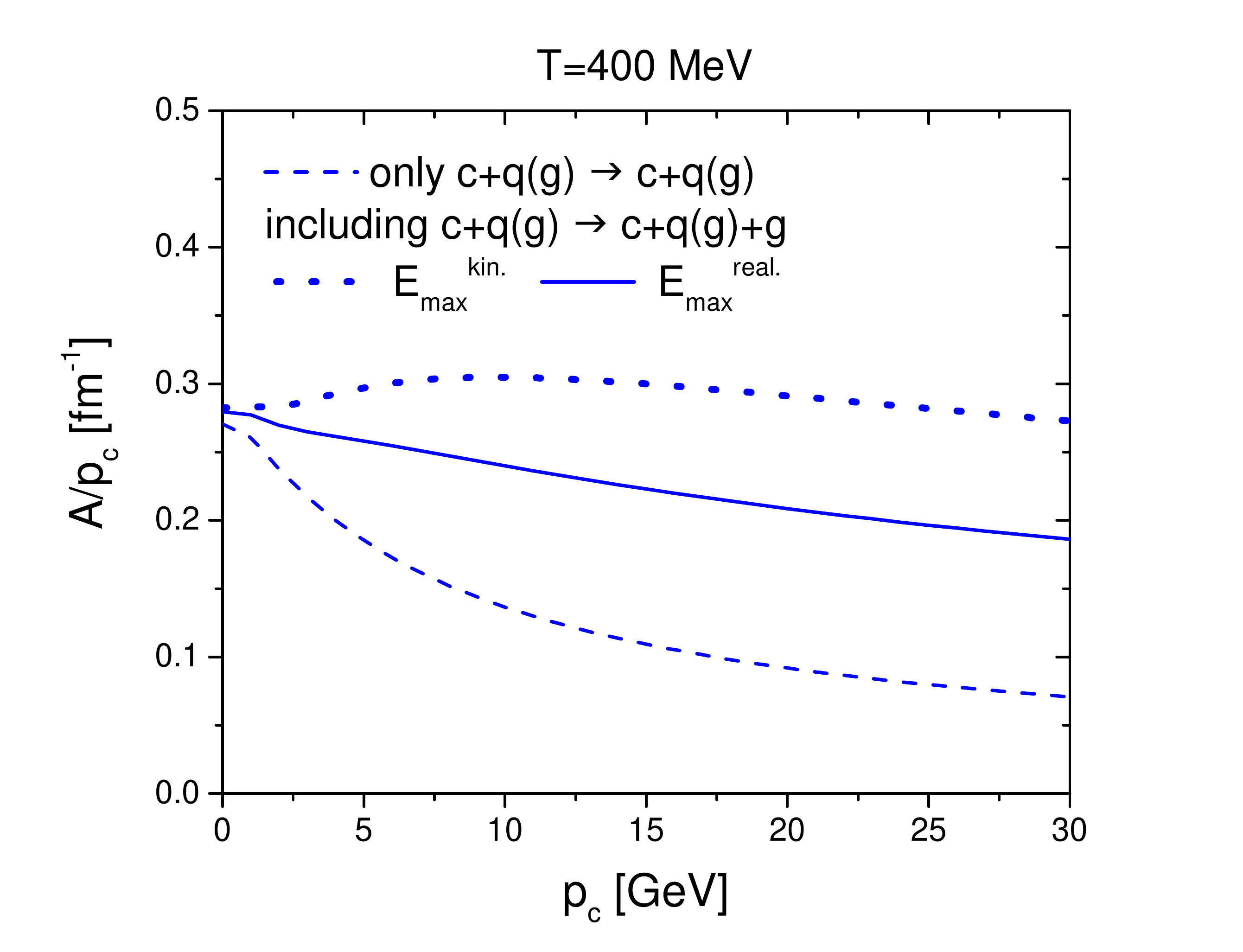}}
\centerline{
\includegraphics[width=8.6 cm]{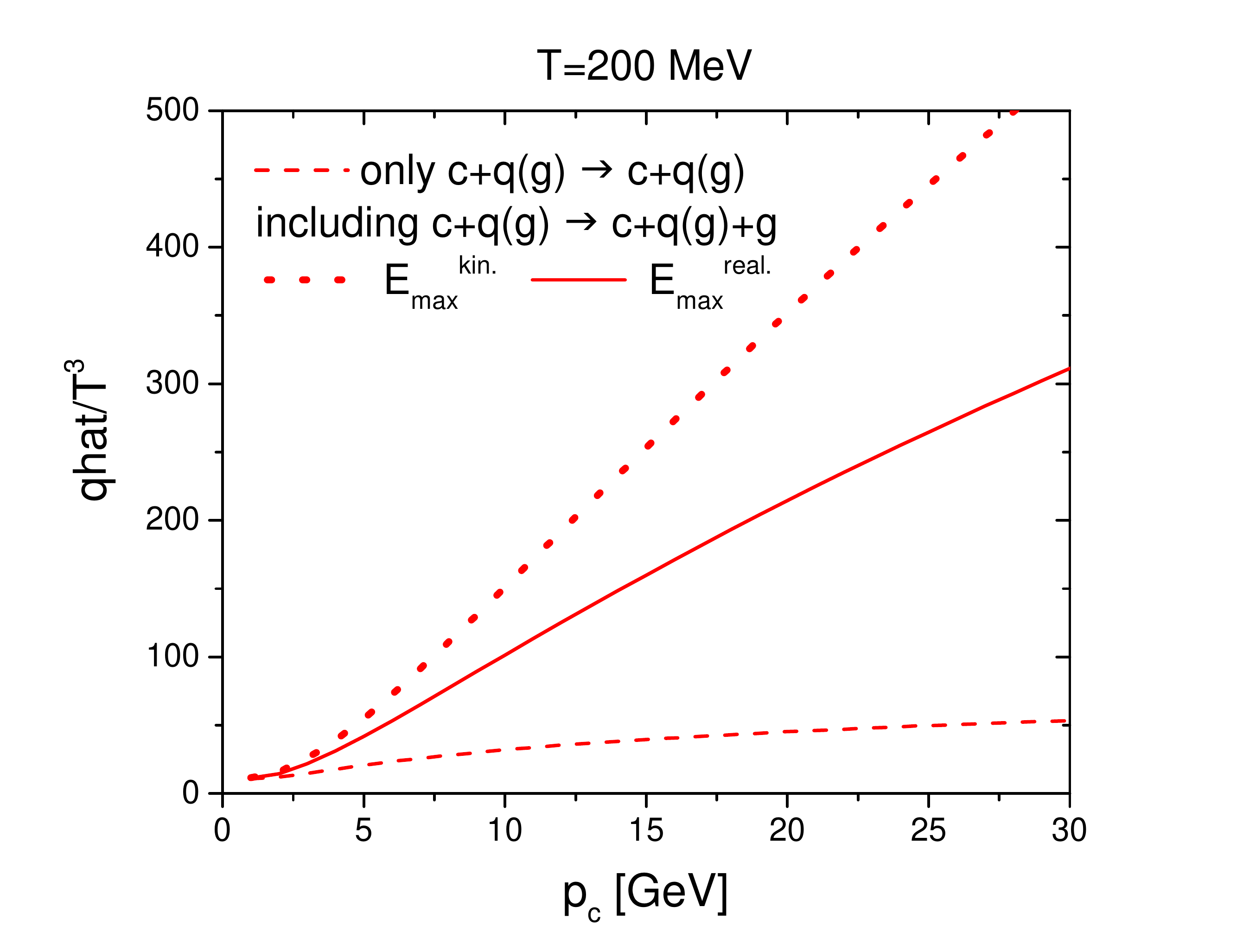}
\includegraphics[width=8.6 cm]{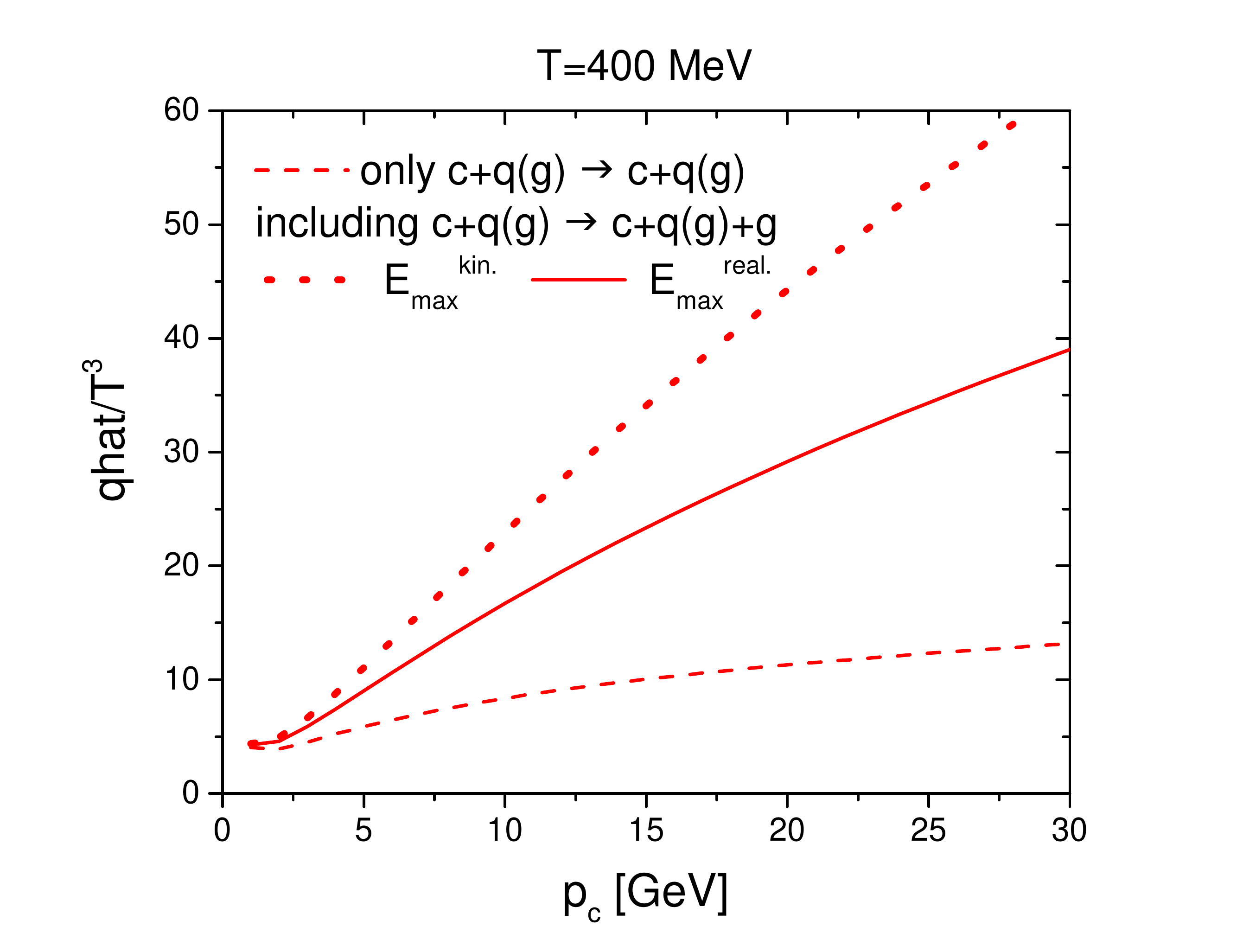}}
\caption{(Color online) (upper) Drag coefficients and (lower) $\hat{q}/T^3$ of the charm quark at T=200 MeV (left) and 400 MeV (right) without and with $c+q(g)\rightarrow c+q(g)$ and $c+q(g)\rightarrow c+q(g)+g$ for the kinematic and realistic upper limits of soft gluon energy, which are, respectively, displayed by dashed, dotted and solid lines.}
\label{transcoeff}
\end{figure*}

Figure~\ref{transcoeff} shows the drag coefficients and $\hat{q}/T^3$ of the charm quark without and with soft gluon emission for the kinematic and realistic upper limits of soft gluon energy at T=200 MeV and 400 MeV.
One can see that 2-to-3 scattering hardly changes the transport coefficients at low momentum, because the scattering energy of a slow charm quark with a thermal parton is not large enough to produce a massive gluon.
As a result, the spatial diffusion coefficient of a heavy quark in the QGP, which is obtained from the drag coefficient in the static limit of charm quark, is not much affected by the 2-to-3 scattering:
\begin{eqnarray}
D_s=\lim_{p_c\rightarrow 0} \frac{T}{m_c (A/p_c)},
\end{eqnarray}
where $A$ is the drag coefficient shown in the upper panels of Fig.~\ref{transcoeff}.

\begin{figure}[th!]
\centerline{
\includegraphics[width=8.6 cm]{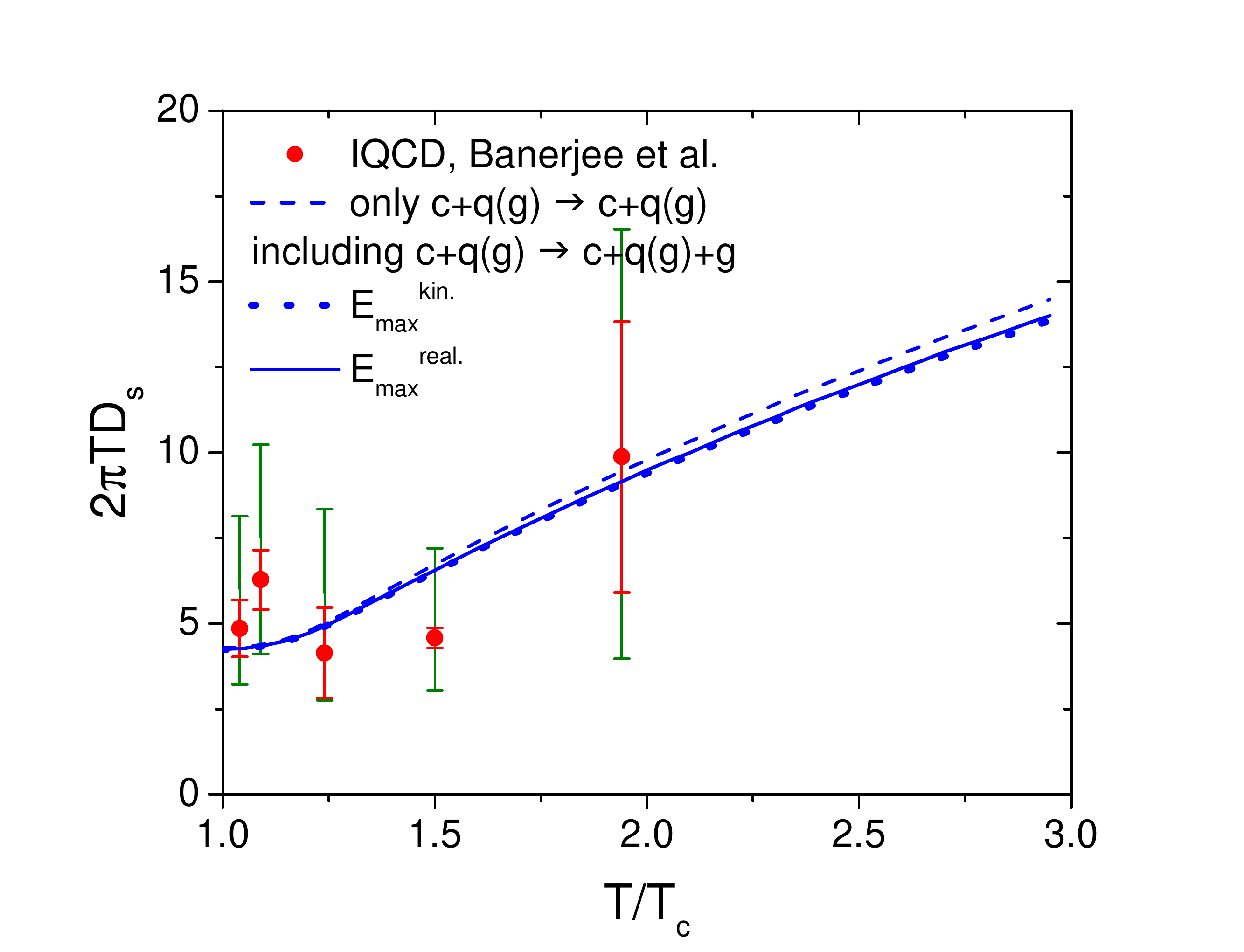}}
\caption{(Color online) spatial diffusion coefficient of charm quark for only elastic scattering and both elastic and radiative scatterings in comparison with lattice data~\cite{Banerjee:2011ra}.}
\label{ds}
\end{figure}

The spatial diffusion coefficient is presently available in lattice calculations~\cite{Banerjee:2011ra} and it is well reproduced within the DQPM~\cite{Berrehrah:2014kba,Song:2016lfv}.
Therefore, this good reproduction will not change even after including the 2-to-3 processes in the DQPM, as shown in Fig.~\ref{ds}.

On the other hand, the radiative energy loss enhances the drag and $\hat{q}$ of the charm quark at large momentum, which is not consistent with the PHSD results presented in Refs.~\cite{Song:2015sfa,Song:2015ykw}, because the $R_{\rm AA}$ of D mesons in heavy-ion collisions at RHIC and LHC are well reproduced only with elastic scattering.
Recently we have found in Ref.~\cite{Grishmanovskii:2022tpb} that the strong coupling $\alpha_s (T)$ - extracted from the lattice EoS - seems to overestimate jet quenching and the mixture of $\alpha_s (T)$ and a constant strong coupling may be more realistic, because an energetic parton is far off thermal equilibrium.
This kind of change of the strong coupling at large momentum will help the PHSD to reproduce experimental data with the radiative energy loss, too.

We note that the transport coefficients from the kinematic upper limit and those from the realistic upper limits are not much different from each other, while the difference of the scattering cross sections is huge in Fig.~\ref{sigma}.
The reason is that the two cross sections mostly differ near the forward scattering, which is not so effective to increase the transport coefficients.

\begin{figure*}[th!]
\centerline{
\includegraphics[width=8.6 cm]{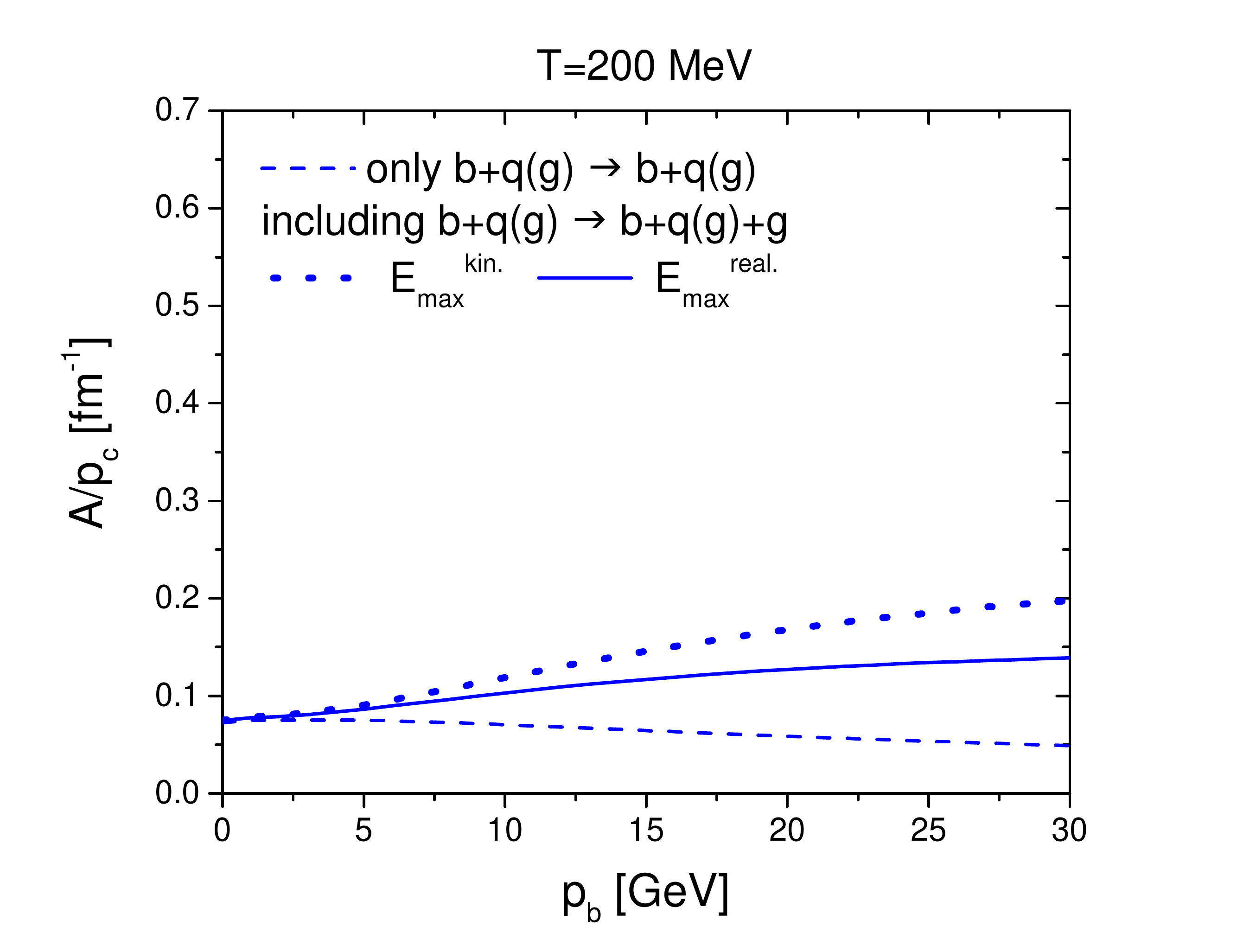}
\includegraphics[width=8.6 cm]{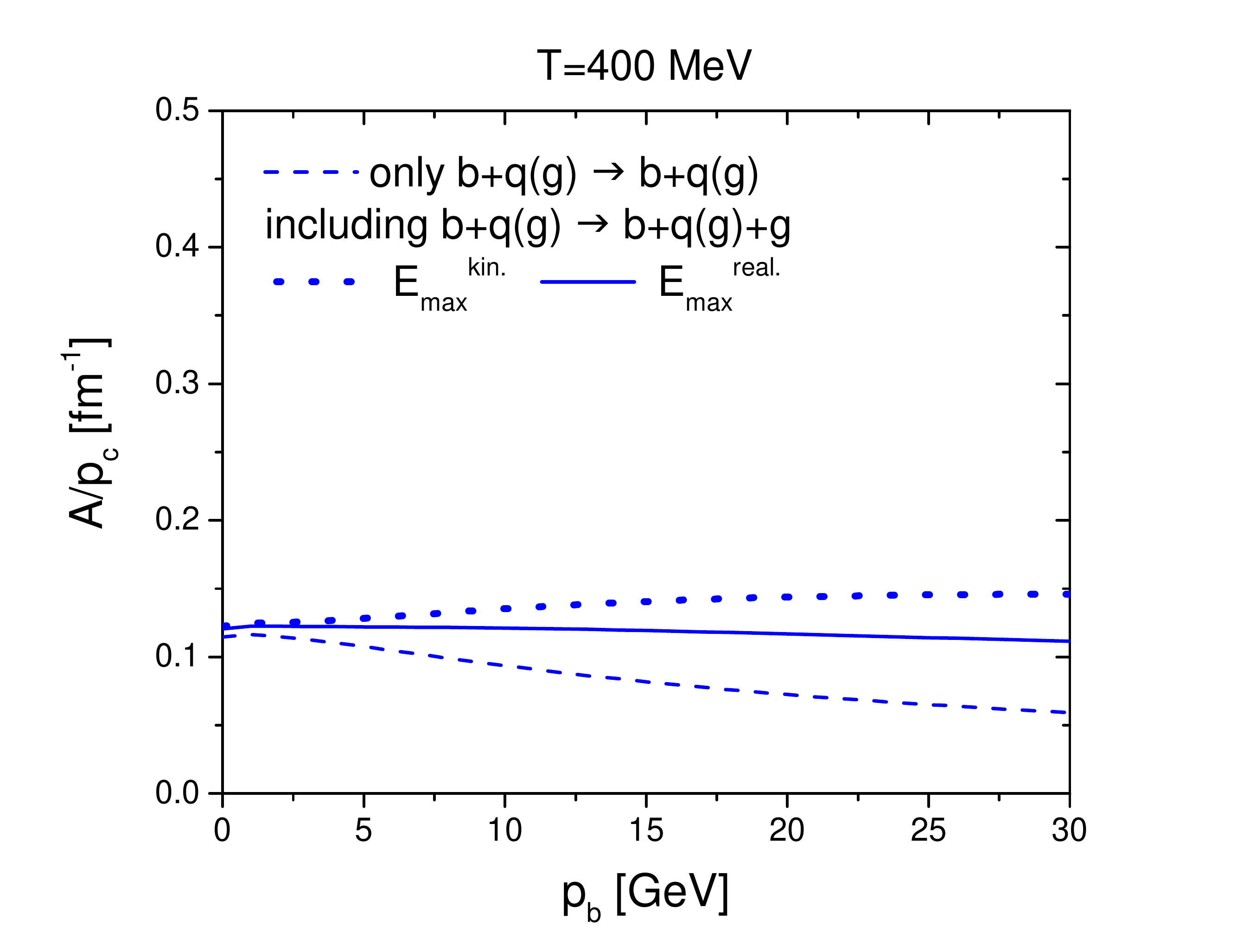}}
\centerline{
\includegraphics[width=8.6 cm]{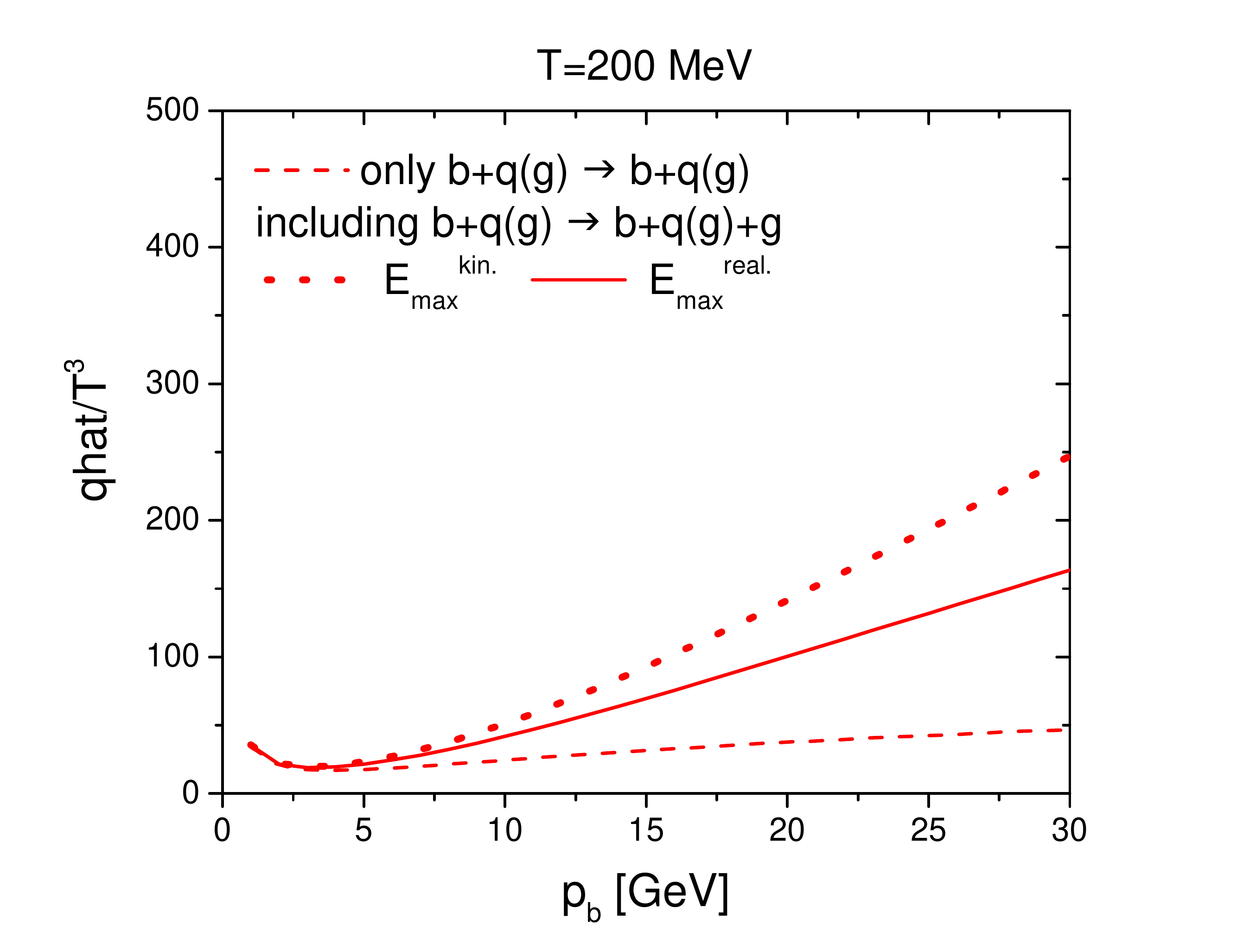}
\includegraphics[width=8.6 cm]{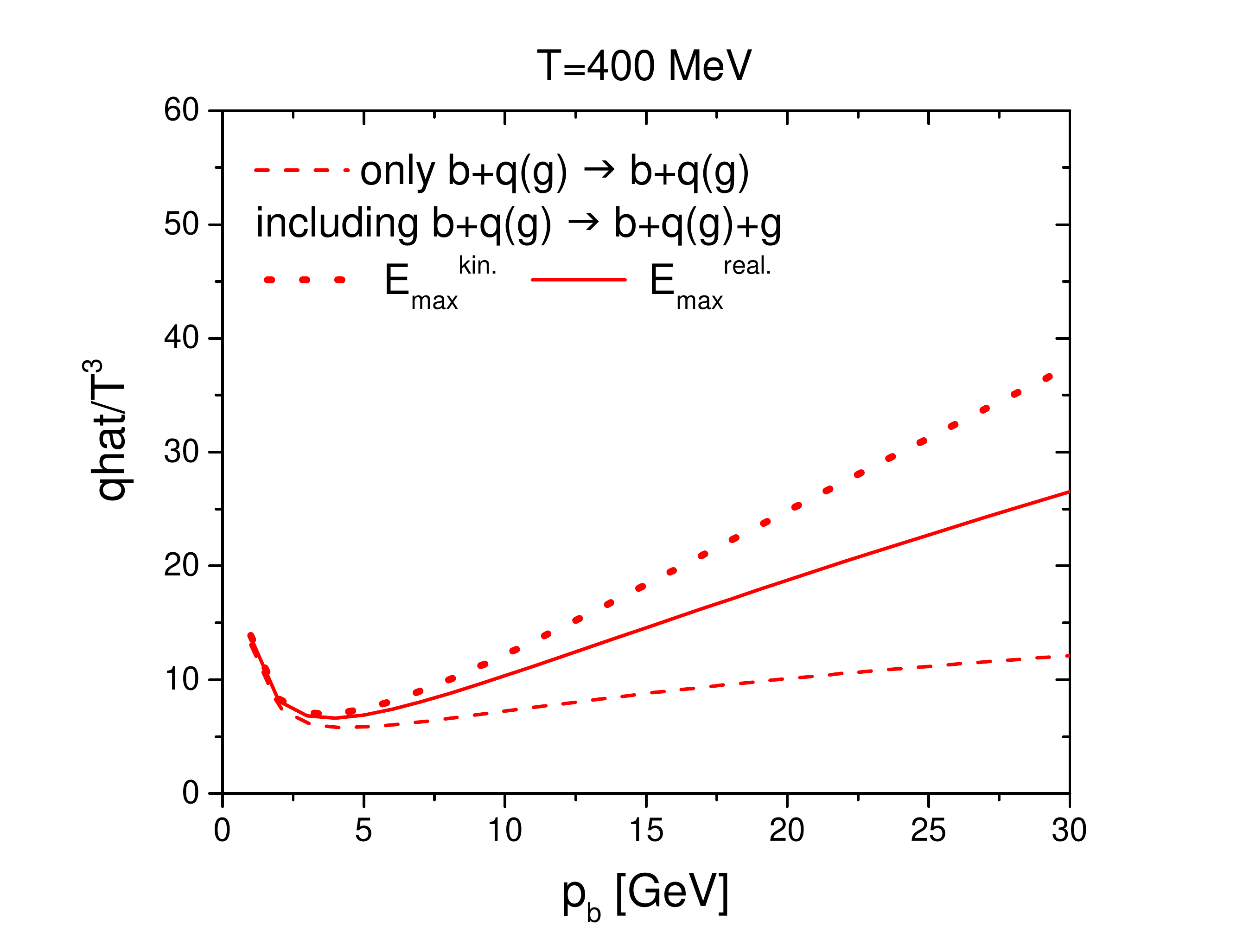}}
\caption{(Color online) same as figure~\ref{transcoeff} but for bottom quark.}
\label{transcoeff-bottom}
\end{figure*}

Figure~\ref{transcoeff-bottom} is same as the figure~\ref{transcoeff} except that charm quark is substituted with bottom quark whose mass is taken to be 4.8 GeV.
Compared with figure~\ref{transcoeff}, the transport coefficients of bottom quark seem to be less affected by the gluon emission.
In fact, it is attributed to the large mass of bottom quark which is about three times larger than charm quark mass. 
A fair comparison between charm and bottom will be made when $p_T$ of bottom quark is rescaled by three times, because interaction rate is proportional to heavy quark velocity.
We also note that $\hat{q}$ of bottom quark is a bit rising at low momentum, because $\hat{q}$ is defined as $d\langle(\Delta p_T)^2\rangle /d z$ rather than $d\langle(\Delta p_T)^2\rangle /d t$ in Eq.~(\ref{qhat}).
The transport coefficients of bottom quark are in general smaller than those of charm quark, because the differential cross section of bottom quark is more highly peaked in forward direction due to the large mass~\cite{Song:2016rzw}.

\section{Summary}\label{summary}

In this study we have extended the formalism for soft photon emission from the scattering of electric-charged particles to soft gluon emission from the scattering of color-charged particles, i.e. of partons.
We have found that the soft gluon is emitted from incoming or outgoiong partons as the emission of soft photon, and the scattering amplitude with the approximation satisfies the Slavnov-Taylor identities.  
This is so because the Slavnov-Taylor identities are satisfied order by order, if the scattering amplitude is expanded in terms of the emitted gluon energy divided by the energy of the scattering particles.
It enables the factorization of radiative scattering into elastic scattering and gluon/photon emission and guarantees gauge-invariance.

Heavy flavor is an important probe particle searching for the properties of an extremely hot and dense matter produced in ultra-relativistic heavy-ion collisions.
We have applied the formalism for the soft gluon emission to the heavy quark scattering off massive quarks or gluons in the sQGP.
For the soft gluon approximation to be valid, emitted gluon energy is restricted up to the energy-momentum transfer of 2-to-2 scattering.

Comparing the integrated cross sections for elastic scattering and for soft gluon emission, we have found that
the latter is strongly suppressed at low momentum, because the scattering energy is not large enough to produce a massive gluon, and that the scattering for gluon emission is more strongly suppressed with increasing temperature than the elastic scattering, because the former is proportional to $\alpha_s^3 (T)$ while the latter to $\alpha_s^2 (T)$.

The results have been extended to the calculations of the transport coefficients of heavy quarks in the sQGP.
For the same reason as for the cross section the transport coefficients little change at small momentum of heavy quark in spite of incluing gluon emission, which means that the spatial diffusion coefficient of heavy quark is not affected by the radiative scattering.
However, the transport coefficients are enhanced by it with increasing charm quark momentum.
Since the energetic parton is far off thermal equilibrium, it is doubtable that the thermal strong coupling extracted from the lattice EoS, $\alpha_s (T,\mu)$, can be applied to the scattering of the energetic parton.
Instead a perturbative strong coupling or a mixture of them may be more reasonable, which will be our next study.

\section*{Acknowledgements}
The authors acknowledge helpful discussions with E. Bratkovskaya, J. Aichelin and W. Cassing.
We acknowledge support by the Deutsche Forschungsgemeinschaft (DFG, German Research Foundation) through the grant CRC-TR 211 "Strong-interaction matter under extreme conditions" - project number 315477589 - TRR 211. I.G. also acknowledges support from the "Helmholtz Graduate School for Heavy Ion research". 
The computational resources have been provided by the LOEWE-Center for
Scientific Computing and the "Green Cube" at GSI, Darmstadt.


\hfil\break
\appendix
\bigskip

\section{}\label{qq-app}

We first show the current conservation of Eq.~(\ref{conservation-q}).
From Eqs.~(\ref{quark}) and (\ref{m_mu}) we get
\begin{eqnarray}
q_\mu M_{2\rightarrow 2+g}^{\mu, a, kl;ij}&=&g\bigg\{M_{2q\rightarrow 2q}^{kl;mj}T^a_{mi}+M_{2q\rightarrow 2q}^{kl;im}T^a_{mj} \nonumber\\
&&-T^a_{km}M_{2q\rightarrow 2q}^{ml;ij}-T^a_{lm}M_{2q\rightarrow 2q}^{km;ij}\bigg\}\nonumber\\
&\sim& T^b_{km}T^b_{lj}T^a_{mi}
+T^b_{ki}T^b_{lm}T^a_{mj}\nonumber\\
&&-T^a_{km}T^b_{mi}T^b_{lj}-T^a_{lm}T^b_{ki}T^b_{mj}.
\end{eqnarray}

Using Eq.~(\ref{colorq}) for $M_{2q\rightarrow 2q}^{kl;ij}$,
\begin{eqnarray}
q_\mu M_{2\rightarrow 2+g}^{\mu, a, kl;ij}
\sim \bigg(T^a_{li}\delta_{kj}-\frac{1}{N_c}T^a_{ki}\delta_{lj}\bigg)~~~~~~~~~~~~~~~\nonumber\\
+\bigg(T^a_{kj}\delta_{il}-\frac{1}{N_c}T^a_{lj}\delta_{ki}\bigg)-\bigg(T^a_{kj}\delta_{il}-\frac{1}{N_c}T^a_{ki}\delta_{lj}\bigg)\nonumber\\
-\bigg(T^a_{li}\delta_{kj}-\frac{1}{N_c}T^a_{lj}\delta_{ki}\bigg)\equiv T_1+T_2-T_3-T_4=0.~~
\end{eqnarray}

Now we turn to the transition amplitude squared, for which the following combinations of color structure are needed:
\begin{eqnarray}
|T_1|^2=|T_2|^2=|T_3|^2=|T_4|^2
=\frac{N_c^2-1}{4}C_F,\nonumber\\
T_1T_2^*=T_3T_4^*
=-\frac{1}{2}C_F,\nonumber\\
T_1T_3^*=T_2T_4^*
=-\frac{1}{4} C_F,\nonumber\\
T_1T_4^*=T_2T_3^*
=\frac{N_c^2-2}{4} C_F,
\label{color2}
\end{eqnarray}
where 
$(T^{a}_{ij})^*=T^{a}_{ji}$ and $C_F=(N_c^2-1)/(2N_c)$.

Taking into account the color factors in Eq.~(\ref{color2}), the transition amplitude squared turns to
\begin{eqnarray}
|M_{2q\rightarrow 2q+g}|^2=-\frac{g^2}{2N_c}\bigg[
(N_c^2-1)\bigg\{
\frac{m_1^2}{(p_1\cdot q)^2}+\frac{m_2^2}{(p_2\cdot q)^2}\nonumber\\
+\frac{m_3^2}{(p_3\cdot q)^2}+\frac{m_4^2}{(p_4\cdot q)^2}\bigg\}-4\frac{p_1\cdot p_2}{(p_1\cdot q)(p_2\cdot q)}\nonumber\\
-4\frac{p_3\cdot p_4}{(p_3\cdot q)(p_4\cdot q)}+2\frac{p_1\cdot p_3}{(p_1\cdot q)(p_3\cdot q)}+2\frac{p_2\cdot p_4}{(p_2\cdot q)(p_4\cdot q)}\nonumber\\
-2(N_c^2-2)\bigg\{\frac{p_1\cdot p_4}{(p_1\cdot q)(p_4\cdot q)}+\frac{p_2\cdot p_3}{(p_2\cdot q)(p_3\cdot q)}\bigg\}\bigg]\nonumber\\
\times|M_{2q\rightarrow 2q}|^2,~~~~~
\end{eqnarray}
where $N_c C_F/2$ is absorbed into $|M_{2q\rightarrow 2q}|^2$.

\section{}\label{qg-app}

Substituting $M_{q+g\rightarrow q+g}^{jb;ia}$ with $if^{abe}T_{ji}^e$ in Eq.~(\ref{gluon1}), one can show the current conservation as follows:
\begin{eqnarray}
q_\mu M_{q+g\rightarrow q+g+g}^{\mu,jbc;ia}(p_1,p_2;p_3,p_4,q)\sim if^{abe}T_{jm}^e T^c_{mi}\nonumber\\
- if^{abe}T^c_{jm}T_{mi}^e -f^{adc}f^{dbe}T_{ji}^e - f^{bdc}f^{ade}T_{ji}^e\nonumber\\
=-f^{abe}f^{ecd}T^d_{ji}-f^{adc}f^{dbe}T_{ji}^e - f^{bdc}f^{ade}T_{ji}^e\nonumber\\
=(-f^{abd}f^{ced} - f^{cbd}f^{ead}-f^{ebd}f^{acd})T_{ji}^e=0,
\end{eqnarray}
which needs the cyclic property of $f^{abc}$ and $[T^e,T^c]_{ji}=if^{ecd}T_{ji}^d$~\cite{Muta:1987mz}.

Now we turn to the scattering amplitude squared.
Considering only color factors, 
\begin{eqnarray}
&&|M_{q+g\rightarrow q+g+g}^{jbc;ia}|^2(p_1,p_2;p_3,p_4,q)~~~~~~~~~~~~~~~\nonumber\\
&\sim& -\bigg| if^{abe}(T^eT^c)_{ji}\frac{p_1^\mu}{p_1\cdot q}-if^{abe}(T^cT^e)_{ji}\frac{p_3^\mu}{p_3\cdot q}\nonumber\\
&&-f^{adc}f^{dbe}T^e_{ji}\frac{p_2^\mu}{p_2\cdot q}
-f^{bdc}f^{ade}T^e_{ji}\frac{p_4^\mu}{p_4\cdot q}\bigg|^2\nonumber\\
&=&\bigg[-\frac{N_c^2-1}{2N_c}\bigg(\frac{m^2}{(p_1\cdot q)^2}+\frac{m^2}{(p_3\cdot q)^2}\bigg)\nonumber\\
&&-\frac{1}{N_c}\frac{p_1\cdot p_3}{(p_1\cdot q)(p_3\cdot q)}+\frac{N_c}{2}\bigg(\frac{p_2\cdot p_4}{(p_2\cdot q)(p_4\cdot q)}\nonumber\\
&&+\frac{p_1\cdot p_2}{(p_1\cdot q)(p_2\cdot q)}+\frac{p_3\cdot p_4}{(p_3\cdot q)(p_4\cdot q)}
+\frac{p_1\cdot p_4}{(p_1\cdot q)(p_4\cdot q)}\nonumber\\
&&+\frac{p_2\cdot p_3}{(p_2\cdot q)(p_3\cdot q)}\bigg)\bigg]\times N_c {\rm Tr}[T^eT^e],
\end{eqnarray}
for which the followings are useful:
\begin{eqnarray}
f^{ade}f^{bef}f^{cfd}&=&\frac{N_c}{2}f^{abc},\nonumber\\
f^{abc}T^bT^c&=&\frac{i}{2}N_cT^a,\nonumber\\
T^bT^aT^b&=&-\frac{1}{2N_c}T^a.
\end{eqnarray}

Since the color structure for $q+g\rightarrow q+g$ is given by
\begin{eqnarray}
|M_{q+g\rightarrow q+g}^{jb;ia}|^2\sim |if^{abe}T_{ji}^e|^2=N_c {\rm Tr}[T^eT^e]
\end{eqnarray}
from $f^{abd}f^{abe}=N_c\delta_{de}$, the scattering amplitude squared for $q+g\rightarrow q+g+g$ turns to
\begin{eqnarray}
&&|M_{q+g\rightarrow q+g+g}|^2=-g^2\bigg[\frac{N_c^2-1}{2N_c}\bigg(\frac{m_1^2}{(p_1\cdot q)^2}+\frac{m_3^2}{(p_3\cdot q)^2}\bigg)\nonumber\\
&&+\frac{1}{N_c}\frac{p_1\cdot p_3}{(p_1\cdot q)(p_3\cdot q)}-\frac{N_c}{2}\bigg(\frac{2p_2\cdot p_4}{(p_2\cdot q)(p_4\cdot q)}\nonumber\\
&&+\frac{p_1\cdot p_2}{(p_1\cdot q)(p_2\cdot q)}+\frac{p_3\cdot p_4}{(p_3\cdot q)(p_4\cdot q)}
+\frac{p_1\cdot p_4}{(p_1\cdot q)(p_4\cdot q)}\nonumber\\
&&+\frac{p_2\cdot p_3}{(p_2\cdot q)(p_3\cdot q)}\bigg)\bigg]\times|M_{q+g\rightarrow q+g}|^2.
\end{eqnarray}

\section{} \label{dqpm_param}
Here we recall main properties and parameters of the DQPM.
We fix the strength of the quasiparticle interaction by the adjusting the coupling constant at $\mu_B =0$ so that quasiparticle entropy density reproduces the entropy density $s(T,\mu_\mathrm{B} = 0)$ from the lattice QCD calculations provided by the BMW collaboration~\cite{Borsanyi:2012cr,Borsanyi:2013bia} in the following way \cite{Berrehrah:2016vzw}:
	\begin{equation}
	g^2(T,\mu_\mathrm{B} = 0) = d \cdot \Big[ \left(s(T,0)/s^\mathrm{QCD}_{\mathrm{SB}} \right)^e -1 \Big]^f, 
	\label{coupling_DQPM}
	\end{equation}
where  $s_{\mathrm{SB}}^{\mathrm{QCD}}/T^3 = 19  \pi^2/9 $ is the Stefan-Boltzmann limit of entropy density for massless quarks and gluons and the dimensionless parameters $d = 169.934$, $e = -0.178434$ and $f = 1.14631$.

At finite $\mu_\mathrm{B}$ the $g^2$ is obtained by employing the `scaling hypothesis' introduced in Ref. \cite{Cassing:2007nb}. It assumes that $g^2$ is a function of the ratio of the effective temperature
	 \begin{equation}
	 T^* = \sqrt{T^2+\mu^2_q/\pi^2}
	 \label{eq:tstar}
	 \end{equation}
 (where the quark chemical potential is defined as $\mu_q=\mu_u=\mu_s=\mu_B/3$ ) and the $\mu_\mathrm{B}$-dependent critical temperature $T_c(\mu_\mathrm{B})$ defined as in Ref. \cite{Berrehrah:2016vzw}:
	 	\begin{equation}
	    T_c(\mu_B) = T_c(0) \sqrt{1-\alpha \mu_B^2},
	    \label{eq:dqpmTc}
	\end{equation}
where  $T_c(0)$ is the critical temperature at vanishing chemical potential ($T_c(0) \approx 0.158$ GeV) and $\alpha = 0.974$ GeV$^{-2}$.
Thus, the DQPM effective coupling constant $g_{DQPM}^2(T,\mu_\mathrm{B})$ reads
\begin{align}
g_{DQPM}^2(T,\mu_\mathrm{B}) \equiv \begin{cases}
& \mu_\mathrm{B} = 0: \ g^2(T,\mu_\mathrm{B} = 0)  \\
& \mu_\mathrm{B} > 0: \ g^2(T_{scale}(T,\mu_\mathrm{B}))
	\end{cases}
\end{align}\label{eq:as_DQPM}
with $T_{scale} =T^*/(T_c(\mu_\mathrm{B})/T_c(0))$.

In the DQPM the quasiparticle pole masses are adopted in the form of asymptotic quark or gluon masses respectively $m_\infty \sim m_D/2$ or $\sqrt{2}m_{f}$, where $m_D$ is the HTL Debye mass, and $m_{f}$ is the HTL thermal fermion mass \cite{Bellac:2011kqa,Linnyk:2015rco}:
	\begin{equation}
	 m^2_{g}(T,\mu_\mathrm{B})=C_g \frac{g^2(T,\mu_\mathrm{B})}{6}T^2\left(1+\frac{N_{f}}{2N_{c}}
	+\frac{1}{2}\frac{\sum_{q}\mu^{2}_{q}}{T^2\pi^2}\right)
		  \label{polemassg_dqpm},
		\end{equation}
		\begin{equation}
	  m^2_{q(\bar q)}(T,\mu_\mathrm{B})=C_q \frac{g^2(T,\mu_\mathrm{B})}{4}T^2\left(1+ \frac{\mu^{2}_{q}}{T^2\pi^2}\right)
	  \label{polemassq_dqpm}.
	  \end{equation}
In Eqs. (\ref{polemassg_dqpm}),(\ref{polemassq_dqpm}) $N_{c}=3$ and $N_{f}=3$ denote the number of colors and the number of flavors respectively, $C_q = \dfrac{N_c^2 - 1}{2 N_c} = 4/3$ and $C_g = N_c = 3$ are the QCD color factors for quarks and for gluons, respectively. 
The strange quark has a larger bare mass which enhances its dynamical mass. This essentially suppresses the channel $g \rightarrow s + {\bar s}$ relative to the channel $g \rightarrow u + {\bar u}$ or $d + {\bar d}$ and controls the strangeness ratio in the QGP. Empirically $m_s(T,\mu_B)= m_u(T,\mu_B)+ \Delta m = m_d(T,\mu_B)+ \Delta m$ where $\Delta m$ =30 MeV has been used \cite{Moreau:2019vhw}.
This model parameter has been fixed in an empirical way by comparing to experimental data for strange hadron abundances and the $K^+/\pi^+$ ratio from heavy-ion collisions at relativistic energies obtained within -- the PHSD approach -- a microscopic covariant transport approach. \\

Furthermore, thermal widths in the DQPM are adopted in the following form \cite{Berrehrah:2016vzw,Linnyk:2015rco}:
	\begin{equation}
		\gamma_{j}(T,\mu_\mathrm{B}) = \frac{1}{3} C_j \frac{g^2(T,\mu_\mathrm{B})T}{8\pi}\ln\left(\frac{2c_m}{g^2(T,\mu_\mathrm{B})}+1\right),
	\label{eq:widths}
	\end{equation}
where the parameter $c_m = 14.4$ was fixed in \cite{Cassing:2007nb}, which is related to a magnetic cut-off. Furthermore, we assume that all (anti-)quarks have the same thermal width: $\gamma_{u}=\gamma_{d}=\gamma_{s}$

	
\bibliography{revision}

\end{document}